\documentclass[sigconf,screen]{acmart}
\AtBeginDocument{%
  }

\usepackage{amsmath}
\usepackage{mathtools}
\usepackage{amsthm}
\usepackage{booktabs}
\usepackage{multirow}

\usepackage{tikz}
\usepackage{comment}
\usepackage{microtype,breakcites}
\usepackage{wrapfig,floatflt,float}

\usepackage{url}
\usepackage{graphicx}
\usepackage{multirow}
\usepackage{caption}
\usepackage{subcaption}
\usepackage{framed, tabu,adjustbox}
\usepackage{xcolor}
\usepackage{color}
\usepackage{stmaryrd}

\usepackage{enumitem}
\usepackage{mdframed}
\usepackage{tikz}

\usepackage{ragged2e}

\usepackage{tikz}
\usepackage{wrapfig}

\usepackage{filecontents}

\theoremstyle{plain}

\theoremstyle{definition}

\theoremstyle{remark}

\usepackage{tikz}
\usetikzlibrary{shapes.geometric, arrows.meta, positioning}

\tikzset{
    block/.style={rectangle, rounded corners, draw=black, fill=red!20, thick, minimum height=2em, minimum width=4em, text centered},
    arrow/.style={-{Stealth}, thick},
}

\newcommand{\name}{LRD-MPC}
\newcommand{\cross}{\times}

\newcommand{\para}[1]{\textbf{#1:}}
\newcommand{\pc}[1]{\ensuremath{#1}\text{-PC}}
\newcommand{\pcent}[1]{$#1\%$}
\newcommand{\dimt}[2]{[$#1\cross #2$]}
\newcommand{\dimf}[4]{[$#1\cross #2\cross #3\cross #4$]}

\newcommand{\partyset}{\ensuremath{\mathcal{P}}}
\newcommand{\party}[1]{\textsf{P}_{#1}}

\newcommand{\shr}[2]{\ensuremath{\llbracket #1 \rrbracket_{#2}}}
\newcommand{\as}[2]{[#1]_{#2}}

\newcommand{\mA}{\ensuremath{\mathbf{A}}}
\newcommand{\mB}{\ensuremath{\mathbf{B}}}
\newcommand{\mM}{\ensuremath{\mathbf{M}}}
\newcommand{\mW}{\ensuremath{\mathbf{W}}}
\newcommand{\mU}{\ensuremath{\mathbf{U}}}
\newcommand{\mC}{\ensuremath{\mathbf{C}}}

\newcommand{\mE}{\ensuremath{\mathbf{E}}}
\newcommand{\mX}{\ensuremath{\mathbf{X}}}
\newcommand{\mY}{\ensuremath{\mathbf{Y}}}
\newcommand{\mZ}{\ensuremath{\mathbf{Z}}}

\newcommand{\mS}{\ensuremath{\mathbf{S}}}

\newcommand{\mV}{\ensuremath{\mathbf{V}}}
\newcommand{\mR}{\ensuremath{\mathbf{R}}}

\newcommand{\valf}{\mathsf{f}}

\newcommand{\valx}{\mathsf{x}}
\newcommand{\valy}{\mathsf{y}}

\newcommand{\mmul}{\odot}
\newcommand{\conv}{\oast}

\newcommand{\procn}[1]{\Pi_{\textsf{#1}}}

\newcommand{\MatMul}{\ensuremath{\mathsf{MatMul}}}
\newcommand{\Trunc}{\ensuremath{\mathsf{Trunc}}}

\newcommand{\sizem}{\mathsf{m}}
\newcommand{\sizen}{\mathsf{n}}
\newcommand{\sizeo}{\mathsf{o}}
\newcommand{\sizer}{\mathsf{r}}
\newcommand{\sizeh}{\mathsf{h}}
\newcommand{\sizew}{\mathsf{w}}
\newcommand{\sizei}{\mathsf{i}}
\newcommand{\sizeb}{\mathsf{b}}

\newcounter{itemcount}

\newenvironment{myenumerate}
{\setcounter{itemcount}{0}\begin{list}
{\arabic{itemcount}.}{\usecounter{itemcount} \itemindent=0.1cm
\itemsep=0.0in
\parsep=0.0in
\topsep=2pt
\partopsep=0.0in}}{\end{list}}

\def\cross{\times}

\newcommand{\nZ}{\mathbb{Z}}

\newcommand{\algoHead}[1]{\vspace{0.2em} \underline{\textbf{#1}} \vspace{0.3em}}
\newcommand{\ms}[2]{\mathsf{m}_{#1,#2}}
\newcommand{\lm}[2]{\lambda_{#1}^{#2}}
\newcommand{\aset}[1]{\{#1\}}
\newcommand{\trunc}[1]{\ensuremath{{#1}^{\valf}}}

\newcommand{\band}{5Gbps}
\newcommand{\lat}{35ms}

\newcommand{\relu}{\ensuremath{\mathsf{ReLU}}}
\newcommand{\gcn}{\ensuremath{\mathsf{GCN}}}
\newcommand{\graphconv}{\ensuremath{\mathsf{GraphConvolution}}}
\newcommand{\bert}{BERT\textsubscript{\scriptsize LARGE}}

\begin{document}
\begin{abstract}
Secure Multi-party Computation (MPC) enables a group of untrusted parties to collaboratively compute a function while ensuring that no party learns any information about the inputs. The application of MPC to machine learning (ML) has gained significant attention. ML inference services can now be hosted on multiple cloud virtual machines (VMs), where each VM acts as an MPC party. The model provider can secretly share their model across VMs while the users can secretly share their inputs. In this setting, MPC offers cryptographically robust security guarantees for both model weights and input data; namely, every MPC server only sees random data for both model parameters and inputs on which they operate. However, MPC protocols pay significant computational and communication overheads. Deep neural network-based machine learning (ML) algorithms heavily depend on convolutional and fully connected layers. For performing a matrix multiplication of two matrices of dimensions \dimt{m}{n} and \dimt{n}{o} \pc{n} MPC protocols perform $3*m*n*o$ multiplications and communicate $m*n*o$ elements across parties. 


%


In this paper, we propose leveraging low-rank approximation for the linear layers of ML models, which has been demonstrated to work well in ML algorithms. 
Low-rank decomposition replaces one matrix multiplication between large matrices with two matrix multiplications between smaller matrices. Each matrix multiplication in MPC incurs a round of communication, meaning decomposing one matrix multiplication into two leads to an additional communication round. Second, the added matrix multiplication requires an additional truncation step, which is necessary to maintain precision after each multiplication. Since truncation involves both communication and computation, these overheads can potentially outweigh the benefits of reducing multiplications through low-rank approximation. To address these challenges, we propose two complementary optimizations: truncation skipping and efficient linear layer concatenation. Truncation skipping is designed to eliminate the additional truncation steps introduced by low-rank decomposition, while efficient linear layer concatenation leverages an optimized pipeline to conceal the added round of communication. Together, these techniques address the key overheads of low-rank decomposition in MPC, enhancing overall performance.

Our proposed optimizations are broadly applicable across various MPC protocols. In this paper, we experimentally demonstrate the benefits of \name{} in both \pc{n} and \pc{3} protocols. Our results show that \name{} achieves up to $25\%$ performance improvement in \pc{n} protocols and $33\%$ in \pc{3} protocols compared to the full-rank baseline. Moreover, \name{} improves GPU energy efficiency by up to $52\%$ and reduces offline phase latency by up to $88\%$, maximizing the benefits of low-rank decomposition in MPC ML.

\end{abstract}
\title{\name{}: Efficient MPC Inference through Low-rank Decomposition}

\author{Tingting Tang}
\affiliation{%
  \institution{University of Southern California}
  \city{Los Angeles}
  \state{California}
  \country{USA}}
\email{tangting@usc.edu}

\author{Yongqin Wang}
\affiliation{%
 \institution{University of Southern California}
  \city{Los Angeles}
  \state{California}
  \country{USA}}
\email{yongqin@usc.edu}

\author{Murali Annavaram}
\affiliation{%
 \institution{University of Southern California}
 \city{Los Angeles}
 \state{California}
 \country{USA}}
\email{annavara@usc.edu}

\maketitle

\section{Introduction}
\label{sec:introduction}
ML models are able to accurately perform a variety of tasks, such as image classification and graph network analysis. ML providers are relying on cloud computing platforms to deploy their models for wider use. Users then submit their inputs to the cloud-hosted models for inference. However, both the model weights and user-provided inputs are often sensitive or proprietary. In a cloud environment, these data are vulnerable to a wide range of threats, including compromised operating systems, physical access breaches, side-channel attacks, and other potential vulnerabilities~\cite{snoop1, specre, meltdown, hashemi2022data}.
Multi-party computation (MPC) protocols provide cryptographic proven security in the cloud, enabling both model providers and users to interact securely in an untrusted cloud environment. The model providers can secretly share their model weights across multiple untrusted cloud virtual machines (VMs), while users can also secretly share their inputs. These VMs then perform computations on the encrypted data and communicate with each other as required by the protocol to complete the inference process.

In this paper, we focus on matrix multiplications, a fundamental operation in deep neural network (DNN)-based ML algorithms. All linear layers (convolutional and fully connected layers) in DNNs are implemented as matrix multiplications. As we show in our result section, in MPC-based ML, matrix multiplication is a major contributor to inference runtime (\pcent{36}-\pcent{51}) and energy consumption (\pcent{67}-\pcent{86}), primarily due to increased computation and additional communication rounds. In the MPC setting, compared to plaintext matrix multiplications, MPC servers perform approximately $2\times$ and $3\times$ more multiplications in \pc{3} and \pc{n} protocols, respectively. These computations also require one round of communication per multiplication. Furthermore, since MPC represents operands in fixed-point format, each multiplication is followed by a truncation step to maintain precision (see Section~\ref{subsec:truncation} for details). Truncation itself requires additional computation and communication.  These extra computations and communication overheads make matrix multiplication a leading bottleneck, significantly impacting both inference latency and energy efficiency. Thus, optimizations on linear layers in ML can widen MPC's appeal for secure ML.

%


In this paper, we propose \name{}, which contains a collection of optimizations to enable fast and energy-efficient MPC inference through low-rank decomposition. Low-rank approximation~\cite{pufferfish} has been demonstrated to work well in ML inference, as model weight matrices have been shown to exhibit low rank. By exploiting the low rank property, a single large matrix multiplication can be reduced to performing two smaller-dimension matrix multiplications. However, naively performing low-rank decomposition will not give us the best performance. We draw this conclusion based on the following observations:
 
\para{Added truncation}  When applying low-rank approximation, MPC parties will need to evaluate two matrix multiplications over smaller dimensions rather than one matrix multiplication over large matrices.  Hence, low-rank approximation adds an additional truncation overhead. Even though the matrix dimension is reduced, each truncation for MPC  incurs one round of additional communication between parties. This added communication can be a significant portion of runtime for MPC configurations with restricted network bandwidth and latency demands.

\para{Added rounds of communication} Network latency is a major factor affecting MPC performance. In general, each round of communication involves a fixed network latency followed by the data transmission. In the low-rank approach, the fixed network latency is paid twice for each multiplication and twice for the truncation steps. In our measurements, the network latency is almost 35 milliseconds on unloaded servers communicating between Southern California and Seattle. Thus, even with a reduced matrix dimension with low rank, the added network latency of low rank can exceed the benefits of smaller matrix multiplications. 

\subsection{Our contribution} Based on those key observations, we propose \name{}, a collection of optimizations for faster MPC inference through low-rank decomposition. Our contribution in \name{} includes:

\para{Low-rank decomposition for MPC ML} We apply low-rank decomposition to linear layers in ML models in an MPC setting. Our preliminary evaluations showed that the benefits from reduced matrix dimensions did not lead to the expected reduction in inference latency. Further analysis of the results highlighted the bottlenecks of added computation and communication due to additional truncation and multiplication resulting from low-rank decomposition. 

\para{Truncation skipping} Based on the analysis, we propose truncation skipping in low-rank MPC. When a linear layer in a model is approximated with low-rank decomposition, truncation skipping will omit the truncation steps for the first low-rank matrix multiplication. However, skipping the truncation in one layer must be accounted for during the final truncation step. We propose a modified truncation approach after the second multiplication, where more bits are truncated for the second low-rank linear layers. Our truncation protocol's runtime does not depend on the number of bits truncated. Hence, we effectively eliminate the one truncation step with no impact on latency. Note that ML inference requires limited precision, as has been demonstrated by numerous quantization approaches~\cite{quantization1, quantization2}. The MPC protocols we use in our experiments use 64-bit fixed-point representation with 5 bits of precision, which rarely causes overflows. Hence, skipping a truncation does not negatively impact model accuracy in ML inference.

\para{Linear layer concatenation} Truncation skipping opens up new avenues for improving MPC performance. In particular, two linear layer computations are performed back-to-back, which is an unusual opportunity that is not present in the full-rank models. We exploit the availability of two back-to-back matrix multiplications to overlap computation and communication, hiding the extra communication induced by two multiplications in low-rank  MPC.

\para{Applicability to many MPC protocols} Our techniques above can be applied to many MPC protocols. We demonstrate our optimization for a semi-honest \pc{n} and \pc{3}~\cite{trio} protocols. 

\para{Concrete ML performance benefit} Using a detailed end-to-end implementation of low rank based ML we  demonstrate up to \pcent{33} end-to-end performance improvement of \name{} for the semi-honest \pc{n} and semi-honest \pc{3}~\cite{trio} protocols. In addition to runtime reduction, \name{} lowers energy consumption and offline cost by up to \pcent{52} and \pcent{88}, respectively.

\subsection{Paper organization}
The rest of the paper is organized as follows: Section~\ref{sec:related-works} discusses related works. Section~\ref{sec:background} discusses notations and related background information. Section~\ref{sec:motivation} presents motivation data, and Section~\ref{sec:main} details our contribution and security analysis of our optimization. Section~\ref{sec:experiment} shows the benefits of \name{} in terms of runtime reduction, energy cost reduction, and offline phase cost reduction. Finally, Section~\ref{sec:conclusion} concludes the paper.

\section{Related works}
\label{sec:related-works}
MPC-based privacy-preserving machine learning (PPML) frameworks optimize secure computation via two primary approaches, protocol design and framework implementation. SecureML \cite{secureml} was among the first to enable deep learning training in MPC, utilizing additive secret sharing for linear layers and Yao’s garbled circuits \cite{yao} for non-linear operations, with efficient transformations between these sharing types. Subsequent protocols have refined secret-sharing-based PPML \cite{aby3,astra, fantasticfour, swift,tetrad}. Building on these advancements, Trio \cite{trio} minimizes computational complexity in semi-honest \pc{3} by reducing correlations between party shares, while CompactTag \cite{compacttag} focuses on optimizing Message Authentication Code tag computation overheads for deep neural networks under an actively secure, dishonest-majority model. While we demonstrate our optimizations on semi-honest \pc{n} and Trio’s semi-honest \pc{3} protocol, our techniques are applicable to a broader range of MPC protocols that use secret sharing for linear layer computations.

Beyond protocol design, another research direction leverages software and hardware optimizations, particularly GPU acceleration, to improve MPC-based PPML performance. CrypTen \cite{crypten2020} extends PyTorch \cite{pytorch} to enable efficient \pc{n} computation in the semi-honest setting with GPU support. CryptGPU \cite{cryptgpu} enhances this by implementing both linear and non-linear operations on GPUs via floating-point CUDA \cite{cuda} kernels for fixed-point arithmetic, achieving up to 8$\times$ speed improvements over CPU-based frameworks. Further improvements include Piranha \cite{piranha} which utilizes NVIDIA’s CUTLASS library \cite{cutlass} to provide native integer kernels for fixed-point computations, leading to a 4$\times$ speedup over CryptGPU. PIGEON \cite{pigeon} dynamically switches between arithmetic vectorization and bitslicing for non-linear layers on the CPU while offloading linear layers to the GPU, achieving 1–2 orders of magnitude higher throughput for large ImageNet batch sizes. Rather than developing a new framework, our work optimizes linear layer inference in a way that is portable across MPC-based PPML frameworks. To validate this, we evaluate \name{} on both CrypTen and PIGEON, demonstrating its broad applicability and performance benefits.

While a majority of the overhead in MPC stems from communication during non-linear operations like ReLU and Softmax\cite{wangCharacIspass}, prior research has addressed this by designing MPC-friendly approximations \cite{securenn, cryptflow2, Falcon, mpcformer}, developing model architectures with fewer non-linear layers \cite{deepreduce,snl,kundu}, or optimizing existing MPC algorithms for non-linear operations \cite{hummingbird}. In contrast, our work focuses on reducing the overhead in linear operations, which also significantly impacts inference runtime in MPC.

Low-rank approximation has been widely used to accelerate plaintext ML traning and inference \cite{pufferfish,lowranklth, lora, anotherfish}, leveraging the inherent low-rank structure of model weight matrices \cite{lowrankproperty}. However, its application in MPC-based ML inference remains largely unexplored. To the best of our knowledge, this is the first work to apply low-rank approximation to optimize linear operations in the MPC setting, improving inference efficiency.
\section{Background}
\label{sec:background}
\subsection{Notation}
\label{subsec:notation}
We used the following notations in this paper. 
$\partyset =\{ \party{1}, \party{2}, ..., \party{n} \}$ denotes a set of MPC parties.
%
%
Boldfont capital letters like  $\mX$ are used to denote a matrix.
We will use tuples of smaller letters like \dimt{m}{n} and \dimf{m}{n}{o}{p} to denote the dimension of matrices.
$\mX \mmul \mY$ denotes matrix multiplication between $\mX$ and $\mY$.
$\mX \conv \mY$ denotes 2D convolution between $\mX$ and $\mY$.
We will use doubly square bracket $\shr{\mX}{i}$ to denote the secret share of $\mX$ in $\party{i}$. Depending on the protocols, $\shr{\mX}{i}$ can be a single matrix or a tuple of matrices.
We will use singly square bracket $\as{\mX}{i}$ to denote additive secret sharing, such that $\mX = \sum_{1}^{n} \as{\mX}{i}$. Note that $\shr{\mX}{i}$ and $\as{\mX}{i}$ are different, $\shr{\mX}{i}$ denote generic secret sharing, and $\as{\mX}{i}$ represents a specific secret share matrix $\mX$, based on the MPC protocol used.

\subsection{MPC overview}
\label{subsec:mpcover}
MPC protocols allow a collection of untrusted parties, denoted as $\partyset =\{ \party{1}, \party{2}, ..., \party{n} \}$, to jointly compute a function over inputs. As long as there is a certain number of parties in this $\partyset$ that has not been controlled or corrupted by the adversary, MPC protocols can guarantee input security. In this paper, to demonstrate the performance improvement of \name{}, we will focus on the state-of-art semi-honest \pc{n} protocol using Beaver triples and the latest \pc{3} protocol Trio~\cite{trio}. In the following sections, we will provide a brief overview of those two protocols.

\begin{figure}[h]
    \centering
    \includegraphics[width=4cm]{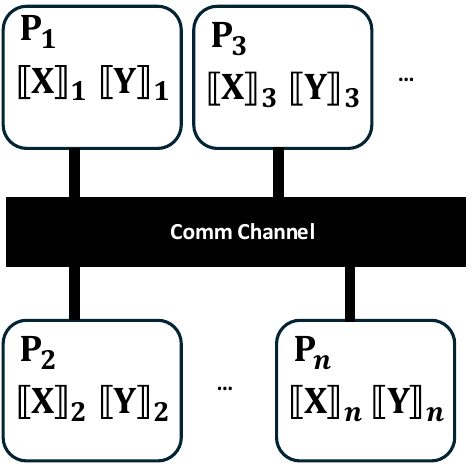}
    \caption{\pc{n} system topology.}
    \label{fig:npc-topo}
\end{figure}

\begin{figure}[h]
    \centering
    \includegraphics[width=4cm]{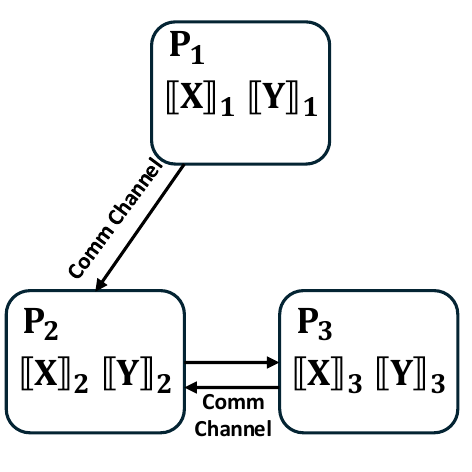}
    \caption{Trio~\cite{trio} system topology.}
    \label{fig:trio-topo}
\end{figure}

\para{Threat model}
\label{subsec:threat}
MPC protocols have different threat models. MPC threat models are characterized by the intention of the adversary and the number of participating parties. MPC's adversaries are categorized into semi-honest (passively malicious) and actively malicious. Both semi-honest and actively malicious adversaries are interested in the user's secret. However, semi-honest adversaries will not modify the final computation results, whereas actively malicious adversaries might. Compared with protocols for semi-honest adversaries, protocols for actively malicious adversaries contain additional mechanisms, such as tag checks, to detect potential errors introduced by adversaries. In this paper, we focus more on semi-honest adversaries because those protocols  require fewer overheads than actively malicious secure MPC protocols.  
Another defining aspect of MPC is the number of participants involved. This paper focuses on semi-honest secure protocols, particularly \pc{n} and \pc{3}, to demonstrate our contributions. In the \pc{n} protocol, security is guaranteed as long as fewer than n parties are compromised by adversaries. In the Trio protocol, security holds as long as no more than one party is compromised. Figure~\ref{fig:npc-topo} illustrates the system topology of \pc{n} protocols, and Figure~\ref{fig:trio-topo} illustrates the system topology for the Trio protocol.

\para{MPC usage procedures}
Three steps are involved in using MPC protocols, and the steps are the following:
\begin{enumerate}
    \item Secretly sharing inputs with all parties.
    \item MPC parties compute secret sharing of outputs.
    \item Retrieving computation results from MPC parties. 
\end{enumerate}
The first step is to secretly distribute inputs to parties while ensuring no parties can gain any information about the original inputs. As shown in Figure~\ref{fig:npc-topo} and~\ref{fig:trio-topo}, all parties will contain $\shr{\mX}{}$ and $\shr{\mY}{}$ after the secret sharing process is done. Then, parties will collectively compute a function on secret shares. During this computation, MPC parties might need to communicate with each other. In Figure~\ref{fig:npc-topo} and~\ref{fig:trio-topo}, \pc{n} parties will need to access broadcast channels, while \pc{3} parties only need to communicate with specific parties during this process. After all parties are done with computation, one can retrieve the results by collecting all the secret shares of the final results from all parties involved. In the following sections, we will describe how \pc{n} and \pc{3} protocols share inputs and compute the final secret-sharing results.

\subsection{\pc{n} protocols}
\label{subsec:npcback}

\para{Secret sharing} The most common secret sharing method in \pc{n} to compute multiplication-related workload is additive secret sharing. For \pc{n} protocols, $\shr{\mX}{i} = \as{\mX}{i}$, and each $\as{\mX}{i}$ appears to be complete random to each party $\party{i}$. 
To reconstruct $\mX$ from $\shr{\mX}{i}$ adversaries needs to collet all $\shr{\mX}{i}$ from all parties. Therefore, security is preserved as long as at least one party remains uncorrupted by adversaries.
Additionally, given two constants $c_1$ and $c_2$, and secret shared matrices $\as{\mX}{i}$ and $\as{\mY}{i}$, $\shr{c_1\mX+c_2\mY}{i} = c_1\as{\mX}{i} + c_2\as{\mY}{i}$, due to linearity of additive sharing. 

\begin{figure}[h]
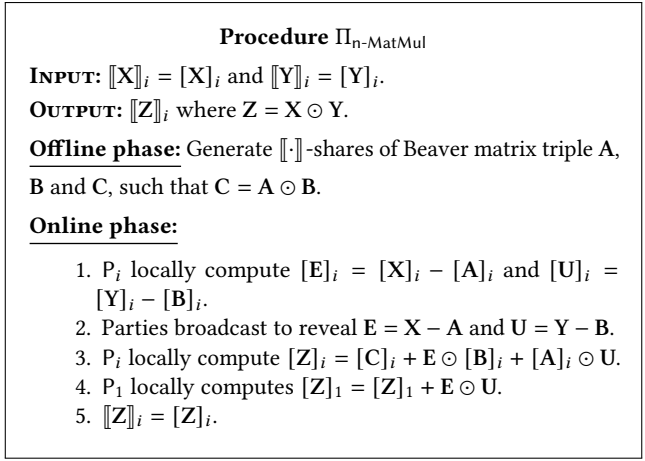

    \begin{framed}
    \centerline{\textbf{Procedure} $\procn{n-\MatMul}$}
    \smallskip

    \begin{flushleft}

    \justify
    
    \textbf{\textsc{Input:}} $\shr{\mX}{i} = \as{\mX}{i}$ and $\shr{\mY}{i} = \as{\mY}{i}$.

    \justify 
    
    \textbf{\textsc{Output:}} $\shr{\mZ}{i}$ where $\mZ = \mX \mmul \mY$.

    \justify 
    
    \algoHead{Offline phase:}
    Generate $\shr{\cdot}{}$-shares of Beaver matrix triple $\mA$, $\mB$ and $\mC$, such that $\mC = \mA \mmul \mB$.
    \justify 
    
    \algoHead{Online phase:} 
    \begin{myenumerate}
        \item $\party{i}$ locally compute $\as{\mE}{i} = \as{\mX}{i} - \as{\mA}{i}$ and $\as{\mU}{i} = \as{\mY}{i} - \as{\mB}{i}$.
        \item Parties broadcast to reveal $\mE = \mX-\mA$ and $\mU = \mY-\mB$.
        \item $\party{i}$ locally compute 
            $\as{\mZ}{i} = \as{\mC}{i} + \mE \mmul \as{\mB}{i} + \as{\mA}{i} \mmul \mU$.
        \item {$\party{1}$ locally computes $\as{\mZ}{1} = \as{\mZ}{1} + \mE \mmul \mU$}.
        \item $\shr{\mZ}{i} = \as{\mZ}{i}$.
    \end{myenumerate}
  \end{flushleft}

  \end{framed}
\caption{Procedure of \pc{n} matrix multiplication.}
\label{fig:nmatmul}
\end{figure}

\para{Matrix Multiplication}
\pc{n} protocols use Beaver triples to perform multiplication among $n$ parties~\cite{crypten2020, wangCharacIspass, mpcpipe}. Beaver triples are secretly shared random matrices $\mA$, $\mB$, and $\mC$, such that $\mC = \mA \mmul \mB$.
Elements in those three matrices are completely random and remain unknown to all MPC parties. Figure~\ref{fig:nmatmul} outlined the detailed procedures to perform matrix multiplication in \pc{n} protocols.
$\procn{n-\MatMul}$ for \pc{n} is divided into two phases: offline and online. Offline phase handles input-independent operations, such as Beaver triple generation. (Figure~\ref{fig:nmatmul} does not specify details about Beaver triple generation as this is a known approach. We refer interested readers to \cite{beavergen} for more details.). Online phase handles input-dependent operations related to $\mX$ and $\mY$.

To perform matrix multiplication, MPC parties first reconstruct $\mX - \mA$ and $\mY-\mB$ as $\mE$ and $\mU$. After reconstructing $\mE$ and $\mU$, each MPC party computes certain local operations to output secret shares of the final results. Adding all $\as{\mZ}{i}s$ together, one will obtain $\mZ = \mX \mmul \mY$. Note that revealing $\mX-\mA$ and $\mY-\mB$ does not reveal information about $\mX$ and $\mY$ since both $\mA$ and $\mB$ are random values that are unknown to all MPC parties. Having covered the fundamentals of \pc{n} protocols, we now present Trio, the latest \pc{3} protocol demonstrating the benefits of \name{}.

\subsection{Trio \pc{3} protocols}
\label{subsec:3pcback}
Trio is introduced in \cite{trio}, and Figure~\ref{fig:trio-topo} shows the system topology of Trio. Unlike \pc{n} protocols, parties in \pc{3} protocols do not need broadcast channels. Instead, $\party{1}$ only needs to send data to $\party{2}$, and does not need to communicate with $\party{3}$ at all. In the meantime, $\party{2}$ receive data from $\party{1}$ and send/receive data to/from $\party{3}$. Thus, a one-direction channel is needed between $\party{1}$ and $\party{2}$, and another bi-directional channel is needed between $\party{2}$ and $\party{3}$. Note that we selected these two diverse protocols to demonstrate our scheme while the protocol design is not our focus.  

\para{Secret sharing} Trio uses a slightly more complex sharing scheme. The following shows how a matrix $\mX$ is shared among those three parties.
\begin{align}
    \shr{\mX}{1} = \aset{\lm{\mX}{2}, \lm{\mX}{3}},\ \  \shr{\mX}{2} = \aset{\ms{\mX}{3}, \lm{\mX}{2}}, \ \ \shr{\mX}{3} = \aset{\ms{\mX}{2}, \lm{\mX}{3}} \nonumber
\end{align}
$\lm{\mX}{2}$ is a random matrix shared between $\party{1}$ and $\party{2}$, and $\lm{\mX}{3}$ is another random matrix shared between $\party{1}$ and $\party{3}$. Moreover, $\ms{\mX}{\{2,3\}}$ is the sum between $\mX$ and $\lm{\mX}{\{2,3\}}$.
\begin{align}
    \ms{\mX}{2} = \mX + \lm{\mX}{2}, \ \ 
    \ms{\mX}{3} = \mX + \lm{\mX}{3} \nonumber
\end{align}
Due to the randomness of $\lm{\mX}{\{2,3\}}$, $\ms{\mX}{\{2,3\}}$ will appear to be completely random to the parties who do not know $\lm{\mX}{\{2,3\}}$. Effectively, $\mX$ is additively shared between $\party{1} \& \party{2}$,  $\party{1} \& \party{3}$, and $\party{2}  \& \party{3}$. Thus, as long as fewer than two of these parties are corrupted, adversaries will not gain any knowledge about the original input $\mX$.

\begin{figure}[h]
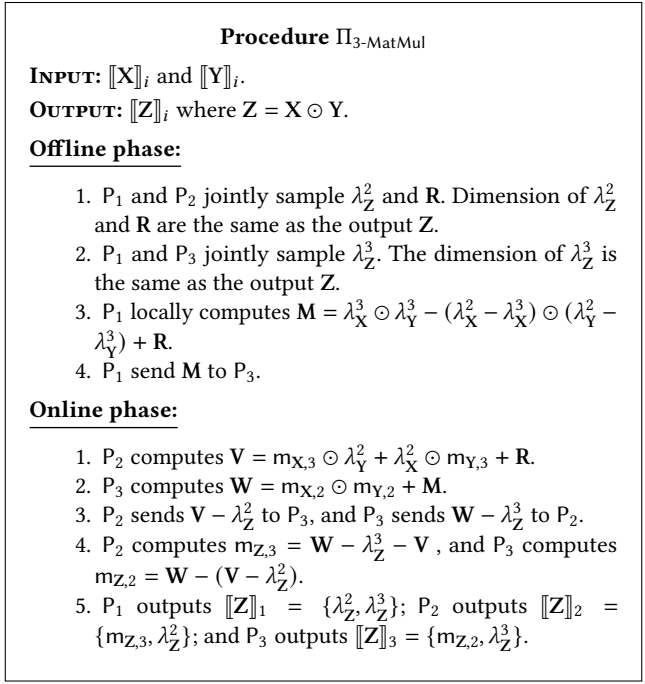

    \begin{framed}
    \centerline{\textbf{Procedure} $\procn{3-\MatMul}$}
    \smallskip

    \begin{flushleft}

    \justify
    
    \textbf{\textsc{Input:}} $\shr{\mX}{i}$ and $\shr{\mY}{i}$.

    \justify 
    
    \textbf{\textsc{Output:}} $\shr{\mZ}{i}$ where $\mZ = \mX \mmul \mY$.

    \justify 
    
    \algoHead{Offline phase:}
    \begin{myenumerate}
        \item $\party{1}$ and $\party{2}$ jointly sample $\lm{\mZ}{2}$ and $\mR$. Dimension of $\lm{\mZ}{2}$ and $\mR$ are the same as the output $\mZ$. 
        \item $\party{1}$ and $\party{3}$ jointly sample $\lm{\mZ}{3}$. The dimension of $\lm{\mZ}{3}$ is the same as the output $\mZ$. 
        \item $\party{1}$ locally computes $\mM = \lm{\mX}{3} \mmul \lm{\mY}{3} - (\lm{\mX}{2}-\lm{\mX}{3}) \mmul (\lm{\mY}{2}-\lm{\mY}{3}) + \mR$.
        \item $\party{1}$ send $\mM$ to $\party{3}$.
    \end{myenumerate}    
    \justify 
    
    \algoHead{Online phase:} 
    \begin{myenumerate}
        \item $\party{2}$ computes $\mV=\ms{\mX}{3}\mmul\lm{\mY}{2} +  \lm{\mX}{2} \mmul \ms{\mY}{3} + \mR$.
        \item $\party{3}$ computes $\mW=\ms{\mX}{2}\mmul\ms{\mY}{2} + \mM$.
        \item $\party{2}$ sends $\mV - \lm{\mZ}{2}$ to $\party{3}$, and $\party{3}$ sends $\mW - \lm{\mZ}{3}$ to $\party{2}$.
        \item $\party{2}$ computes $\ms{\mZ}{3} = \mW - \lm{\mZ}{3} -\mV$ , and $\party{3}$ computes $\ms{\mZ}{2} = \mW - (\mV - \lm{\mZ}{2})$.
        \item $\party{1}$ outputs $\shr{\mZ}{1} = \aset{\lm{\mZ}{2}, \lm{\mZ}{3}}$; $\party{2}$ outputs $\shr{\mZ}{2} = \aset{\ms{\mZ}{3}, \lm{\mZ}{2}}$; and $\party{3}$ outputs $\shr{\mZ}{3} = \aset{\ms{\mZ}{2}, \lm{\mZ}{3}}$.
    \end{myenumerate}
  \end{flushleft}

  \end{framed}
\caption{Procedure of Trio matrix multiplication.}
\label{fig:3matmul}
\end{figure}

\para{Matrix Multiplication} Figure~\ref{fig:3matmul} shows the detailed procedure of matrix multiplication for Trio. Matrix multiplication in Trio also adopts the offline/online computation paradigm. In the offline phase, besides generating shared randomness, $\party{1}$ will compute $\mM$ using local operands and send $\mM$ to $\party{3}$. During the online phases, $\party{2}$ and $\party{3}$ will first compute $\mV$ and $\mW$ and send $\mV$ and $\mW$ to the corresponding recipient. After this communication is done, $\party{2}$ and $\party{3}$ can locally compute $\ms{\mZ}{3}$ and $\ms{\mZ}{2}$. Finally, in the last step of the online phase, each party will output $\shr{\mZ}{i}$.

After executing $\procn{3-\MatMul}$, one can reconstruct $\mZ$ by collecting secret shares of $\mZ$ from two parties in $\aset{\party{1}, \party{2}, \party{3}}$. In MPC, $\procn{3-\MatMul}$ alone is not sufficient for matrix multiplication for MPC ML. A truncation protocol is also needed after every matrix multiplication protocol, and we will discuss this in the next section.

\subsection{MPC Truncation}
\label{subsec:truncation}
\begin{figure}[!t]
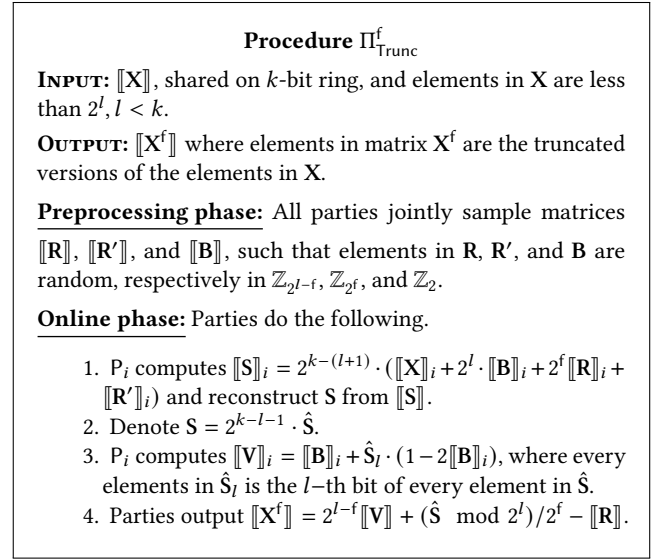

    \begin{framed}
    \centerline{\textbf{Procedure} $\procn{\Trunc}^{\valf}$}
    \smallskip

    \begin{flushleft}

    \justify
    
    \textbf{\textsc{Input:}} $\shr{\mX}{}$, shared on $k$-bit ring, and elements in $\mX$ are less than $2^{l}, l<k$.
    \justify 
    
    \textbf{\textsc{Output:}}  $\shr{\trunc{\mX}}{}$ where elements in matrix $\trunc{\mX}$ are the truncated versions of the elements in $\mX$.

    \justify 
    
    \algoHead{Preprocessing phase:}
    All parties jointly sample matrices $\shr{\mR}{}$,  $\shr{\mR'}{}$, and  $\shr{\mB}{}$, such that elements in $\mR$, $\mR'$, and $\mB$ are random, respectively in $\nZ_{2^{l-\valf}}$,  $\nZ_{2^{\valf}}$, and  $\nZ_{2}$.

    \justify 

    \algoHead{Online phase:} Parties do the following.
    \begin{myenumerate}
        \item $\party{i}$ computes $\shr{\mS}{i}= 2^{k-(l+1)}\cdot (\shr{\mX}{i} + 2^l\cdot \shr{\mB}{i} + 2^{\valf}\shr{\mR}{i} + \shr{\mR'}{i})$ and reconstruct $\mS$ from $\shr{\mS}{}$.
        \item Denote $\mS = 2^{k-l-1}\cdot \hat{\mS}$.
        \item $\party{i}$ computes $\shr{\mV}{i}=\shr{\mB}{i} + \hat{\mS}_{l}\cdot(1-2\shr{\mB}{i})$, where every elements in $\hat{\mS}_{l}$ is the $l-$th bit of every element in $\hat{\mS}$.
        \item Parties output $\shr{\trunc{\mX}}{} = 2^{l-\valf}\shr{\mV}{} + (\hat{\mS} \mod 2^{l})/2^{\valf} - \shr{\mR}{}$.
    \end{myenumerate}

  \end{flushleft}
  \end{framed}
\caption{Procedure for truncating the last $\valf$ bits of each element in a $\shr{\cdot}{}$-shared matrix $\mX$~\cite{newprimitivempc}.}
\label{fig:truncation}
\end{figure}

\para{Fixed-point arithmetic}
The truncation protocol is another key component that MPC parties rely on. To accurately represent real numbers in ML models, parties use fixed-point arithmetic. In this representation, a real number is shared on a $k$-bit ring, with the last $\valf$ bits reserved for precision. For instance, if a value $\valx$ is shared on the ring with $\valf$ precision bits, the value stored in hardware is $\valx \cdot 2^{\valf}$.
When MPC parties multiply their local shares of $\valx$ and $\valy$, the resulting value stored in hardware is $\valx \cdot \valy \cdot 2^{2\valf}$, which corresponds to the real number $\valx \cdot \valy \cdot 2^{\valf}$—not the desired $\valx \cdot \valy$. To correct this, the protocol must remove the trailing $\valf$ bits after each multiplication to ensure accuracy. This removal process is called truncation. Figure~\ref{fig:truncation} details the truncation protocol~\cite{newprimitivempc} we use in this paper. This truncation protocol has been shown to have no errors, even for large ML models~\cite{curl2024}. While recently the security of probabilistic truncation like $\procn{\Trunc}^{\valf}$ was questioned~\cite{prob_trunc_wrong}, \cite{curl2024} has proven the security of probabilistic truncation using a modified ideal functionality. 

Note that $\procn{\Trunc}^{\valf}$, illustrated in Figure~\ref{fig:truncation}, is compatible with both \pc{n} and Trio protocols. To securely execute truncation using $\procn{\Trunc}^{\valf}$, MPC parties simply need to perform the appropriate addition and multiplication operations specific to \pc{n} or Trio when executing $\procn{\Trunc}^{\valf}$. We will see in later sections that this truncation process can pose challenges when low-rank decomposition is applied.

\subsection{Other ML operations}
\label{subsec:endmpc}
Besides matrix multiplication in FC layers, ML models contain other operations like 2D convolution in convolution layers and comparisons in ReLU layers. In this section, we will briefly describe how those layers are computed in MPC ML.

\para{Convolution}
Convolution layers are another type of layer commonly used in ML models and, like FC layers, are also classified as linear layers. In MPC, convolution operations are computed similarly to matrix multiplication. For instance, to perform a 2D convolution between input images $\mX$ and a weight matrix $\mW$, both $\mX$ and $\mW$ are first secret-shared among the MPC parties. The convolution process then follows a similar procedure to that depicted in Figures~\ref{fig:nmatmul} and~\ref{fig:3matmul}, with the key difference being that all matrix multiplications are replaced by their corresponding convolution operations.

\para{Comparison}
Comparison operations are integral to the non-linear layers of ML models. Commonly used layers such as ReLU, MaxPooling, and Softmax rely on these operations. For instance, ReLU compares inputs against zero, while Softmax and MaxPooling involve finding the maximum value among a set of inputs. In MPC, comparison operations are computationally expensive, requiring multiple rounds of communication, which significantly increases the overall runtime of non-linear layers.
Since \name{} focuses on optimizing the linear layers in MPC, we do not delve into the detailed implementation of non-linear layers. For readers interested in these aspects, we  refer to ~\cite{crypten2020, wangCharacIspass}.

\subsection{Low-rank decomposition}
\label{subsec:lowrankback}
In this section, we will discuss how to apply low-rank decomposition to matrix multiplication and convolution layers in ML models.

\para{Low rank matrix multiplication}
For ML models, inference for FC layer is to perform a matrix multiplication between an input matrix $\mX$ (such as a batch of images) and a static weight $\mW$. The static weight in ML models has been shown to satisfy low-rank properties, such that we can decompose a larger matrix $\mW$ into two smaller matrices $\mU$ and $\mV$. $\mW$'s dimension is \dimt{\sizen}{\sizeo} while $\mU$ and $\mV$'s dimension will be \dimt{\sizen}{\sizer} and \dimt{\sizer}{\sizeo}, where $\sizer$ is the estimated rank of $\mW$ and $\mU\mmul\mV \approx \mW$. Assuming the full-rank FC layer is to compute matrix multiplication between input matrix $\mX$, whose dimension is \dimt{\sizem}{\sizen}, low-rank decomposition is shown below:
\begin{align}
    \mX \mmul \mW \approx \mX \mmul \mU \mmul \mV 
    \label{eq:lowconv}
\end{align}
When evaluating the full-rank FC layer, the total number of multiplication is $\sizem\cdot\sizen\cdot\sizeo$, while the total number of multiplications for the low-rank FC layer is $\sizem\cdot\sizen\cdot\sizer + \sizem\cdot\sizer\cdot\sizeo$. As long as $\sizer < \sizen\cdot\sizeo/(\sizen+\sizeo)$, the total number of multiplications for the low-rank FC is fewer than that for full-rank FC. The ratio of $\sizer/min(\sizen,\sizeo)$ is referred to as the rank ratio.

\para{Low rank convolution} Similarly, full-rank convolution layer evaluate this convolution $\mX\conv\mW$ during inference, where $\mX$ of dimension \dimf{\sizeb}{\sizeh}{\sizew}{\sizei} is the input to the layer and $\mW$ of dimension \dimf{3}{3}{\sizei}{\sizeo} is the static weight. Likewise, we will decompose this $\mW$ into two smaller matrices as shown below:
\begin{align}
    \mX \conv \mW \approx \mX \conv \mU \conv \mV 
    \label{eq:lowmm}
\end{align}
Dimensions of $\mU$ and $\mV$ are \dimf{3}{3}{\sizei}{\sizer} and \dimf{1}{1}{\sizer}{\sizeo}, respectively. This low-rank decomposition will also lead to a reduction in computation complexity. Similarly, the ratio of $\sizer/\sizeo$ is referred to as the rank ratio.

Applying low-rank approximations to plaintext models is relatively straightforward, often yielding performance speedups with minimal effort. However, extending these techniques to MPC-based ML introduces additional costs and overheads. We will delve into these challenges and their implications in the following sections.

\section{Motivation}
\label{sec:motivation}
Before introducing \name{}, we first present motivating data on \pc{n} ML, highlighting the importance of optimizing linear layers in MPC-based ML. Linear layers remain a primary contributor to latency and energy consumption in MPC systems, underscoring the need for targeted optimization.

\para{Inference runtime}  
Although activation functions (ReLU, Softmax, and MaxPool) generally still take longer to evaluate in MPC than linear layers~\cite{wangCharacIspass}, recent advancements in MPC frameworks ~\cite{cryptgpu, crypten2020, mpcpipe, piranha, pigeon} and optimizations to activation functions ~\cite{mpcformer, cicra, reluskip, relubits} have reduced their relative impact. Thus, convolution and FC layers now constitute a substantial portion of ML inference.
The first column (Runtime) in Table~\ref{tab:motivation} highlights the percentage of runtime contribution of linear layers during the online phases (Section ~\ref{subsec:npcback}) of \pc{n} ML inference across three representative ML workloads—vision, language, and graph models— demonstrating that linear layers can account for up to \pcent{51.6} of the total runtime. 
 

\begin{table}[h]
    \centering
    \caption{Linear layer runtime (\%) and power (\%) in MPC inference online phase (5Gbps WAN) for \pc{n}.}
    \begin{tabular}{ccc}
        \toprule
        \textbf{Protocol} & Runtime  &  Power \\
        \midrule
        VGG   & 42.8\%  & 72.4\% \\
        \midrule
        BERT\textsubscript{\scriptsize LARGE}  & 51.6\% & 67.8\%\\
        \midrule
        GCN & 36.1\% & 85.3\% \\
        \bottomrule
    \end{tabular}
    \label{tab:motivation}
\end{table}

\para{Power consumption}
In MPC ML, linear layers dominate energy consumption. While activation functions may have longer runtimes due to communication overhead, their energy impact is minimal. This is because activation functions in MPC primarily involve waiting for communication to complete, with only lightweight bitwise operations like XOR and AND, which consume far less energy on cloud servers than convolutional and fully connected layers. Thus, as shown in the second column (Power) in Table~\ref{tab:motivation}, for \pc{n} protocols, linear layers—driven by computationally intensive convolution and FC operations—consume between \pcent{68} and \pcent{86} of GPU energy. 

Given the impact of linear layers on inference runtime and power consumption, optimizing them is crucial to pushing the MPC's efficiency boundaries further. Low-rank approximation has proven effective in plaintext ML, reducing both runtime and energy costs in linear layers~\cite{pufferfish, anotherfish}. However, applying low-rank decomposition in MPC is not straightforward. MPC-specific challenges, such as communication overhead and truncation procedures, limit the potential benefits of low-rank approximation. In the next section, we examine these challenges in detail and present our solutions to overcome them.

\section{\name{}}
\label{sec:main}
In this section, we will discuss challenges for low-rank MPC ML and how we address them.

\begin{figure}[htp]
    \centering
    \begin{subfigure}[t]{0.45\columnwidth} 
        \includegraphics[width=\linewidth]{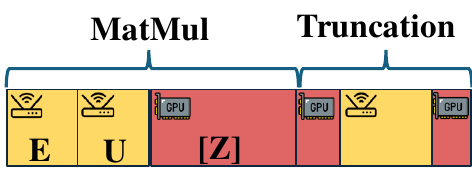} 
        \caption{\pc{n}.}
        \label{fig:npcfflow}
    \end{subfigure}
    \hspace{1em}
    \begin{subfigure}[t]{0.45\columnwidth} 
        \includegraphics[width=\linewidth]{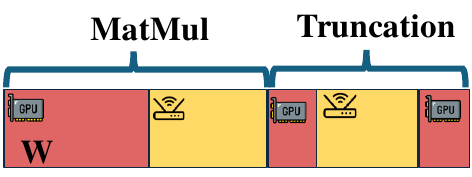} 
        \caption{$\party{3}$ in Trio.}
        \label{fig:3pcfflow}
    \end{subfigure}

    \caption{Computation and communication workflow for \pc{n} and Trio matrix multiplication; boxes labeled with a router icon denote communication runtime, and boxes labeled with a GPU icon denote computation runtime.}
    \label{fig:fflow}
\end{figure}

\subsection{Low rank decomposition}
\label{subsec:lowrank}
Equations~\ref{eq:lowmmss} and~\ref{eq:lowconvss} demonstrate how matrix multiplication and convolution operations are adapted using low-rank decomposition in MPC.
\begin{align}
    \shr{\mX}{} \mmul \shr{\mW}{} \rightarrow \shr{\mX}{} \mmul \shr{\mU}{} \mmul \shr{\mV}{}  \label{eq:lowmmss} \\
    \shr{\mX}{} \conv \shr{\mW}{} \rightarrow \shr{\mX}{} \conv \shr{\mU}{} \conv \shr{\mV}{} \label{eq:lowconvss}
\end{align}

Similar to plaintext ML inference, applying low-rank decomposition to linear layers in MPC reduces the number of multiplications required. 
The reduction in the number of multiplications by low-rank decomposition for linear layers significantly benefits the offline phase of MPC ML. 
For instance, in \pc{n} protocols, a Beaver triple must be generated for each multiplication in the online phase, and generating these triples is the primary overhead~\cite{asplos23character} for \pc{n} protocols. In \pc{3} protocols, $\party{1}$ can spend less time computing $\mM$ for $\party{3}$ (as illustrated in Figure~\ref{fig:3matmul}). However, the advantages of low-rank decomposition for the online phase are less clear-cut. The additional communication rounds and truncation steps can diminish its ability to accelerate the online phase.
\begin{figure}[htp]
    \centering
    \begin{subfigure}[t]{0.49\columnwidth} 
        \centering
        \includegraphics[width=\linewidth]{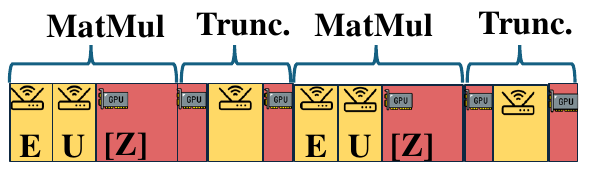} 
        \caption{\pc{n}.}
        \label{fig:npclflow}
    \end{subfigure}
    \hspace{0.3em}
    \begin{subfigure}[t]{0.49\columnwidth} 
        \centering
        \includegraphics[width=\linewidth]{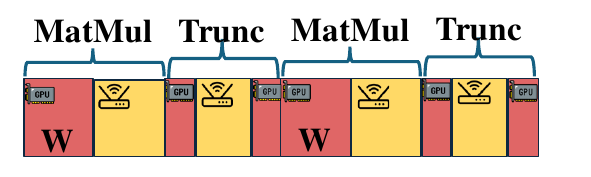} 
        \caption{$\party{3}$ in Trio.}
        \label{fig:3pclflow}
    \end{subfigure}

    \caption{Computation and communication workflow of low-rank matrix multiplication for \pc{n} and Trio protocol.}
    \label{fig:lflow}
\end{figure}

\para{Computation and communication workflow} Figure~\ref{fig:fflow} illustrates the communication and computation workflow of the online phase for full-rank matrix multiplication in \pc{n} (Figure~\ref{fig:npcfflow}) and Trio (Figure~\ref{fig:3pcfflow}) protocols. In Figure~\ref{fig:npcfflow}, parties begin by reconstructing $\mE$ and $\mU$ through broadcasting. Once broadcasting is complete, the parties compute $\as{\mZ}{}$ and subsequently execute the truncation protocol to remove the last $\valf$ bits of $\as{\mZ}{}$, which introduces an additional round of communication. In Figure~\ref{fig:3pcfflow}, $\party{3}$ will compute $\mW$ and then send a masked $\mW$ to $\party{2}$, followed by truncation protocol.
After applying low-rank decomposition to matrix multiplication, the computation and communication workflow of the matrix multiplication is modified as depicted in Figure~\ref{fig:lflow}. For both protocols, a single matrix multiplication with a large matrix is transformed into two matrix multiplications with smaller matrices due to the application of low-rank decomposition. As a result, in both Figure~\ref{fig:npclflow} and Figure~\ref{fig:3pclflow}, we observe two separate matrix multiplications. In MPC ML, each matrix multiplication is immediately followed by a truncation step, meaning both protocols require two truncations with low-rank decomposition.

\para{Added communication} Compared with the full-rank matrix multiplication, low rank decomposition incurs two additional rounds of communication for both \pc{n} protocol and Trio protocols. Those two added communication rounds come from \textbf{1) added matrix multiplication and 2) added truncation}.

Since low-rank decomposition converts one matrix multiplication into two matrix multiplications, and in the MPC setting, each matrix multiplication will incur one round of communication (for the \pc{n} protocol, $\mE$ and $\mU$ can be sent back-to-back, resulting in one round of communication). The other added communication is incurred by truncation. Due to the usage of fixed-point arithmetic, after every multiplication, MPC parties will jointly execute truncation protocols (Figure~\ref{fig:truncation}) to discard trailing $\valf$ bits of the product. Since the truncation protocol will incur one round of communication when using low-rank decomposition for matrix multiplication, additional truncation steps incur one more round of communication. Those added communications, due to additional matrix multiplication and additional truncation protocol, can hinder low-rank decomposition from achieving its full potential for the online phase. For example, when evaluating the low-rank Transformers in BERT models, the reduction in computation time roughly accounts for \pcent{11.4}. However, the added communication mentioned above is able to add \pcent{3.4} communication runtime cost, offsetting the benefits of low-rank to \pcent{7}. The added communication overhead associated with low-rank decomposition prevents it from fully realizing its potential in the MPC setting. 

\name{} incorporates two techniques designed to optimize performance: 1) truncation skipping and 2) efficient layer concatenation. These techniques work synergistically with low-rank decomposition to mitigate communication costs and maximize its overall effectiveness. In the following sections, we will delve into the details of these techniques.
\begin{figure}[htp]
    \centering
    \begin{subfigure}[t]{0.45\columnwidth} 
        \centering
        \includegraphics[width=\linewidth]{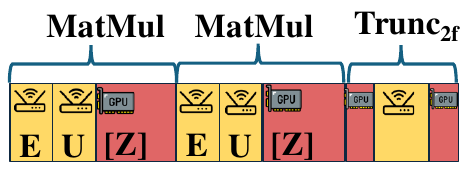} 
        \caption{\pc{n}.}
        \label{fig:npctsflow}
    \end{subfigure}
    \hspace{1em}
    \begin{subfigure}[t]{0.45\columnwidth} 
        \centering        \includegraphics[width=\linewidth]{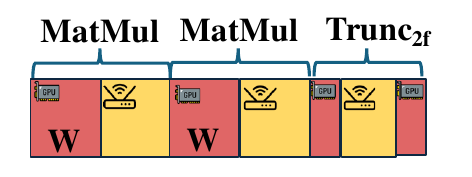} 
        \caption{$\party{3}$ in Trio.}
        \label{fig:3pctsflow}
    \end{subfigure}

    \caption{Computation and communication workflow of low-rank matrix multiplication for \pc{n} and Trio protocol with truncation skipping.}
    \label{fig:tsflow}
\end{figure}

\subsection{Truncation skipping}
In the previous example, we observed that applying low-rank decomposition to linear layers introduces additional truncation steps due to the extra matrix multiplication. To address this issue, we implement a technique called truncation skipping. This approach bypasses the truncation step after the first matrix multiplication and instead adjusts the truncation after the second multiplication to remove $2\valf$ bits from the final product. Specifically, rather than performing $\procn{\Trunc}^{\valf}\{\procn{\Trunc}^{\valf}(\mX\mmul\mU) \mmul \mV\}$, the MPC parties compute $\procn{\Trunc}^{2\valf}(\mX\mmul\mU \mmul \mV)$. This optimization effectively reduces the number of truncation steps while preserving the correctness of the result. Figure~\ref{fig:npctsflow} illustrates the computation and communication workflow after applying truncation skipping.

\para{Correctness} Our truncation skipping technique ensures the correctness of computation by applying a single truncation step that removes $2\valf$ bits at the end. Without any intermediate truncation operations, the actual value stored in computer hardware after evaluating $\shr{\mX}{} \mmul \shr{\mU}{} \mmul \shr{\mV}{}$ is $\shr{\mX}{} \mmul \shr{\mU}{} \mmul \shr{\mV}{} \cdot 2^{3\valf}$. This occurs because multiplying three inputs together accumulates $3\valf$ precision bits in the final product. By performing the $\procn{\Trunc}^{2\valf}$ step at the end, the trailing $2\valf$ bits are removed, leaving the final value in hardware as $\shr{\mX}{} \mmul \shr{\mU}{} \mmul \shr{\mV}{} \cdot 2^{\valf}$. This result correctly represents the intended outcome of $\shr{\mX}{} \mmul \shr{\mU}{} \mmul \shr{\mV}{}$, ensuring that the expected result of low-rank matrix multiplication is achieved without introducing unnecessary truncation overhead during intermediate steps.

\para{Costs of $\procn{\Trunc}^{\valf}$ and $\procn{\Trunc}^{2\valf}$} 
Both the computation and communication costs associated with $\procn{\Trunc}^{2\valf}$ and $\procn{\Trunc}^{\valf}$ at the end of low-rank matrix multiplication are identical, as truncation costs do not scale with the number of truncated bits. As shown in Figure~\ref{fig:truncation}, the online phase involves reconstructing the $\shr{\cdot}{}$-ed $\mS$, which requires one round of broadcasting $\shr{\mS}{}$. The amount of data broadcasted remains constant regardless of the number of bits being truncated. Similarly, the value of $\valf$ does not affect the other computations performed during this online phase.
For the offline phase, the total number of secretly shared bits that need to be generated is $l-\valf+\valf+1$, where $l-\valf$ bits are used for $\mR$, $\valf$ bits for $\mR'$, and one bit for $\mB$. This number is unaffected by the specific number of bits being truncated.  Therefore, by skipping the intermediate truncation step, we effectively remove the associated costs without increasing the cost of the final truncation. This optimization allows us to avoid unnecessary intermediate operations while maintaining overall efficiency.

\para{Impact on model accuracy}  Truncation skipping can potentially impact model accuracy. In MPC, inputs are shared on a $k$-bit ring, where $\valf$ bits are allocated for fractional precision and $k-\valf$ bits for integer representation. Using truncation skipping, more bits are effectively utilized for fractions, leaving fewer bits for integer representation (for the second matrix multiplication). This adjustment may lead to precision loss and could degrade the accuracy of ML models. In \name{}, we address this issue by experimentally evaluating the impact of truncation skipping on model accuracy (Section~\ref{subsec:configuration}). To ensure consistent evaluation while balancing performance and accuracy, we conservatively enable truncation skipping for all low-rank layers across all models.


Through truncation skipping, we successfully eliminate most of the additional costs introduced by the extra truncation steps required due to the additional matrix multiplications. However, this is not the extent of \name{}'s contributions. As observed earlier, the additional matrix multiplications also result in increased rounds of communication, further impacting efficiency. In the next subsection, we will explore how \name{} addresses this issue through an effective linear layer concatenation, which reduces communication overhead and further enhances performance.

\subsection{Efficient linear layer concatenation}
The next technique employed in \name{} is referred to as efficient linear layer concatenation. This method effectively mitigates the additional round of communication introduced by the extra matrix multiplication, seamlessly hiding the overhead into the overall computation to enhance efficiency. As illustrated in Figure~\ref{fig:tsflow}, after applying truncation skipping, MPC parties execute two matrix multiplication protocols back-to-back for both protocols. This setup enables the design of innovative pipeline techniques, allowing us to establish a streamlined communication and computation pipeline. These pipelines effectively hide the extra round of communication caused by the additional matrix multiplication, further optimizing the performance of the system. In this section, we will go into the detail of \pc{n} and \pc{3} protocols to showcase how efficient linear layer concatenation works for both protocols.

\begin{figure}
    \centering
    \begin{subfigure}[t]{0.45\columnwidth}
        \includegraphics[width=\linewidth]{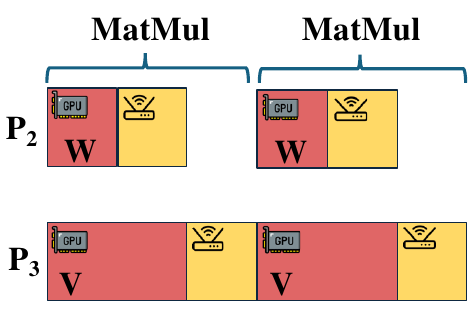}
        \caption{without efficient layer concatenation.}
        \label{fig:trio-flow-mm}
    \end{subfigure}
    \hspace{1em}
    \begin{subfigure}[t]{0.45\columnwidth}
        \includegraphics[width=\linewidth]{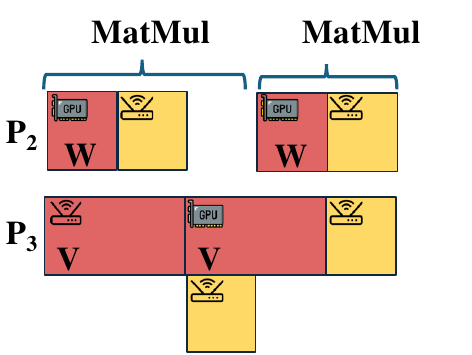}
    \caption{with efficient layer concatenation.}
    \label{fig:trio-flow-pipe}
    \end{subfigure}
    \caption{Trio online phase matrix multiplication computation and communication workflow.}
    \label{fig:fflow}
\end{figure}

\para{Trio}
Let's first see how efficient linear layer concatenation works for the Trio protocol. In the previous section, for simplicity, to showcase the effect of low-rank decomposition and truncation skipping, we only show the computation and communication workflow for $\party{3}$ in Figure~\ref{fig:3pcfflow},~\ref{fig:3pclflow} and~\ref{fig:3pctsflow}. However, that is not the full picture of the Trio protocol. During the online phase, $\party{2}$ and $\party{3}$ will compute $\mV$ and $\mW$ respectively, before sending data to $\party{3}$ and $\party{2}$. 
Figure~\ref{fig:trio-flow-mm} illustrates the computation and communication workflow for $\party{2}$ and $\party{3}$ when low-rank decomposition is applied. As shown in Figure~\ref{fig:3matmul}, $\party{2}$ must perform twice the amount of computation compared to $\party{3}$ before transmitting data to $\party{3}$. This additional workload explains why $\party{2}$'s computation block is significantly longer than $\party{3}$'s. More importantly, $\party{3}$ must wait to receive data from $\party{2}$ before proceeding to the next layer, resulting in periods where $\party{3}$ is entirely idle, waiting for $\party{2}$ to complete its computations and transmission. 

However, the use of low-rank decomposition introduces an interesting synergy between two linear layers, as MPC parties evaluate two matrix multiplications back-to-back. Since the computation in $\party{2}$ is long, before $\party{2}$ finishes computing, it already receives the necessary data from $\party{3}$ to proceed with the second layer. This enables $\party{2}$ to immediately begin processing the next layer as soon as it completes the current one, effectively overlapping its communication with the computation for the subsequent layer. Figure~\ref{fig:trio-flow-pipe} illustrates this pipelined workflow, where part of $\party{2}$'s communication is concealed beneath its computation for the next layer. This overlap reduces the overall latency of executing two matrix multiplications, improving efficiency and minimizing idle periods in the workflow. 

This efficient concatenation is feasible only when two linear layers are executed consecutively. Without low-rank decomposition, each linear layer is immediately followed by a truncation protocol, which lacks sufficient computational workload for $\party{2}$ to overlap its communication with computation. As a result, the communication and computation phases remain sequential rather than overlapping, preventing the latency optimizations achieved through the pipelined execution enabled by \name{}. Besides Trio, other \pc{3} like~\cite{aby3, astra} can also benefit from this efficient concatenation.

\begin{figure}[h]
    \centering
    \includegraphics[width=0.5\linewidth]{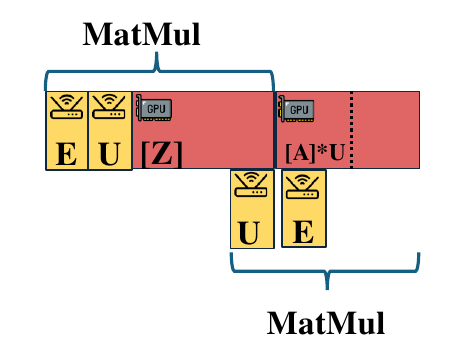}
    \caption{\pc{n} online phase matrix multiplication computation and communication workflow with efficient layer concatenation.}
    \label{fig:nmatmul-pipe}
\end{figure}

\para{\pc{n}} 
In \name{}, we introduce pipelining for linear layers to enable efficient layer concatenation in \pc{n} protocols. As illustrated in Figure~\ref{fig:npctsflow}, parties in the \pc{n} protocol must first broadcast to reconstruct $\mU$ and $\mE$ before proceeding to compute $\shr{\mZ}{}$. However, in ML workloads, one of the inputs for matrix multiplication is typically the model’s static weight. This creates an opportunity to optimize because the broadcasting of $\mU$ for the second layer, derived from static weights, is independent of $\mB$ (prepared during the offline phase). Taking advantage of this lack of dependency, we overlap the broadcasting of $\mU$ for the second matrix multiplication with the computation of $\shr{\mZ}{}$ for the first layer. As shown in Figure~\ref{fig:nmatmul-pipe}, the transmission of $\mU$ is effectively hidden under the computation of $\shr{\mZ}{}$ in the first layer, reducing idle time and improving the overall efficiency of consecutive matrix multiplications in \pc{n} protocols.

In addition to hiding the transmission of $\mU$, we can also conceal the broadcasting of $\mE$ by overlapping it with the local computation of the current layer. Recall that the computation of $\shr{\mZ}{}$ consists of two components: $\mE \mmul \as{\mB}{i}$ and $\as{\mA}{i} \mmul \mU$. Notably, the calculation of $\as{\mA}{i} \mmul \mU$ does not depend on $\mE$. Since $\mU$ for the second linear layer is already received before the first layer's computation is complete, MPC parties can immediately begin computing $\as{\mA}{i} \mmul \mU$ for the second layer as soon as they finish computing the first layer's $\shr{\mZ}{}$, while broadcasting $\mE$ in parallel. As depicted in Figure~\ref{fig:nmatmul-pipe}, this approach effectively hides the transmission of $\mE$ beneath the computation of $\as{\mA}{i} \mmul \mU$ for the second layer, further reducing communication overhead and improving the efficiency of the low-rank MPC models.

In \name{}, we integrate low-rank decomposition into MPC ML models to achieve significant overall performance improvements. Additionally, we enhance the effectiveness of low-rank decomposition by incorporating truncation skipping and efficient linear layer concatenation techniques. These optimizations maximize the performance gains from low-rank approximation, ensuring its full potential is realized in improving computational and communication efficiency. In the next subsection, we will discuss the security of the proposed optimizations.

\subsection{Security analysis}
\para{Low-rank decomposition} MPC ML models continue to preserve security even with the application of low-rank decomposition. Low-rank decomposition transforms a single matrix multiplication involving a large matrix into two matrix multiplications with smaller matrices. Consequently, MPC parties execute $\procn{n-\MatMul}$ and $\procn{3-\MatMul}$ twice for \pc{n} and \pc{3} protocols, respectively. Since  both $\procn{n-\MatMul}$ and $\procn{3-\MatMul}$ are secure by design, ML models employing low-rank decomposition maintain the same security level as their full-rank counterparts. However, one aspect that MPC protocols cannot conceal is the dimension of the matrices being processed. As a result, adversaries can infer the rank chosen for different low-rank layers based on matrix dimensions. Despite this, the protection provided by secret sharing ensures that adversaries gain no information about the actual elements of the inputs, preserving the confidentiality of the data.

\para{Truncation skipping} Truncation skipping eliminates one truncation step in a low-rank matrix multiplication layer but does not alter any other aspects of the underlying MPC protocols. As a result, the security guarantees remain identical to those of low-rank models without truncation skipping, ensuring no compromise in the overall security level.

\para{Efficient linear layer concatenation} The efficient linear layer concatenation technique in \name{} simply rearranges the order of communication and computation without altering the actual communication content. This means that the adversaries observing the protocol will have the same view as they would in the standard MPC protocol without this concatenation. Since no additional information is leaked, and the communication contents remain unchanged, the security guarantees of the underlying MPC protocol are fully preserved. Therefore, the efficient linear layer concatenation technique does not introduce any new vulnerabilities or weaken the protocol's security.

In conclusion, the techniques introduced in \name{}, including low-rank decomposition, truncation skipping, and efficient linear layer concatenation, preserve the security guarantees of the underlying MPC protocols. Truncation skipping eliminates unnecessary steps without altering protocols, while linear layer concatenation merely reorganizes communication without changing its content, ensuring that adversaries gain no additional information. These techniques maintain the same security level as standard MPC implementations. Having established the security of \name{}, we now transition to the evaluation section, where we demonstrate the practical benefits of these techniques through detailed performance and accuracy analyses.
\section{Evaluation}
\label{sec:experiment}

In this section, we present the empirical performance impact of \name{} on MPC ML workflows. The metrics included in this section are inference runtime reduction (online and offline), power consumption, and impact on model accuracy. 

\subsection{Evaluation environment}
\label{subsec:environ}
\para{Server configurations} We evaluate our design using servers equipped with an AMD EPYC 7502 CPU and an Nvidia Quadro RTX 5000 GPU. Each server is connected via an interconnection bandwidth of \band{}, which is widely available across Amazon AWS nodes~\cite{awsbandwidth}. The network latency used in our evaluation is \lat{}, representing the average latency between northern and southern cities in the United States.

\para{MPC Frameworks} We implement our modifications for the Trio protocol using the PIGEON framework~\cite{pigeon}, while our implementation for the \pc{n} protocol is based on CrypTen~\cite{crypten2020}. To provide a comprehensive analysis, we separately measure the timing for the online and offline phases in both frameworks and demonstrate the performance improvements achieved by \name{} for each phase. The diversity of frameworks and models enables us to demonstrate the efficacy of our approach broadly.

\subsection{Models evaluated}
\label{subsec:models}
We evaluate the performance impact of \name{} on four ML models: VGG19~\cite{vgg}, WideResNet18 (WRN18)~\cite{wideresnet}, Graph Convolutional Network (GCN)~\cite{gcn} and BERT\textsubscript{\scriptsize LARGE}~\cite{bert}. VGG19 and WideResNet18 are widely recognized for their effectiveness in computer vision tasks. GCN is a prominent model designed for node classification in large-scale graph datasets. BERT\textsubscript{\scriptsize LARGE} is a widely used Transformer model designed for natural language processing tasks, excelling in tasks such as text classification. By including these models in our evaluation, we demonstrate \name{}'s versatility and performance benefits across a diverse set of ML workloads.

For readers unfamiliar with GCN, the following outlines the algorithm typically used during GCN inference.
\begin{align}
    \gcn(\mX) = \mA \mmul \relu(\mA \mmul\mX\mmul \mW) \mmul \mW'
\end{align}
$\mX$ represents the input node features to be classified, while $\mA$ denotes the adjacency matrix that encodes the graph's structure. The matrices $\mW$ and $\mW'$ are the weight parameters learned during the training phase and are used for node classification during inference. These components work together to propagate information across the graph and make predictions for each node.

\begin{figure*}[!h]
  \centering
  \begin{subfigure}[tb]{0.22\linewidth}
    \includegraphics[width=\linewidth]{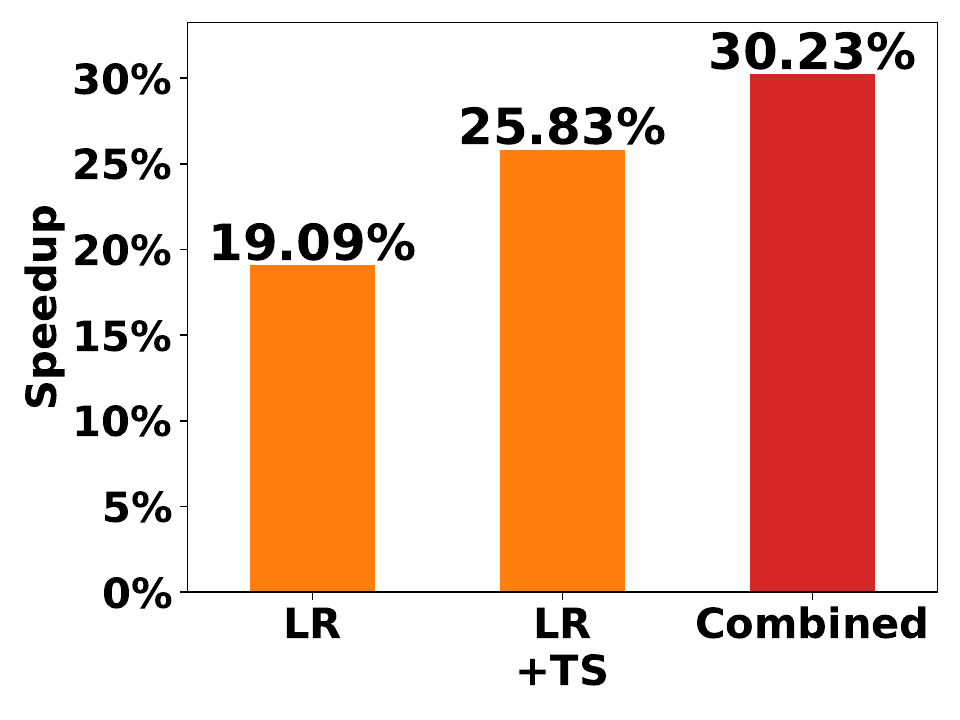}
    \caption{GCN, Cora}
    \label{fig:gcn-3}
  \end{subfigure}
  \begin{subfigure}[tb]{0.22\linewidth}
    \includegraphics[width=\linewidth]{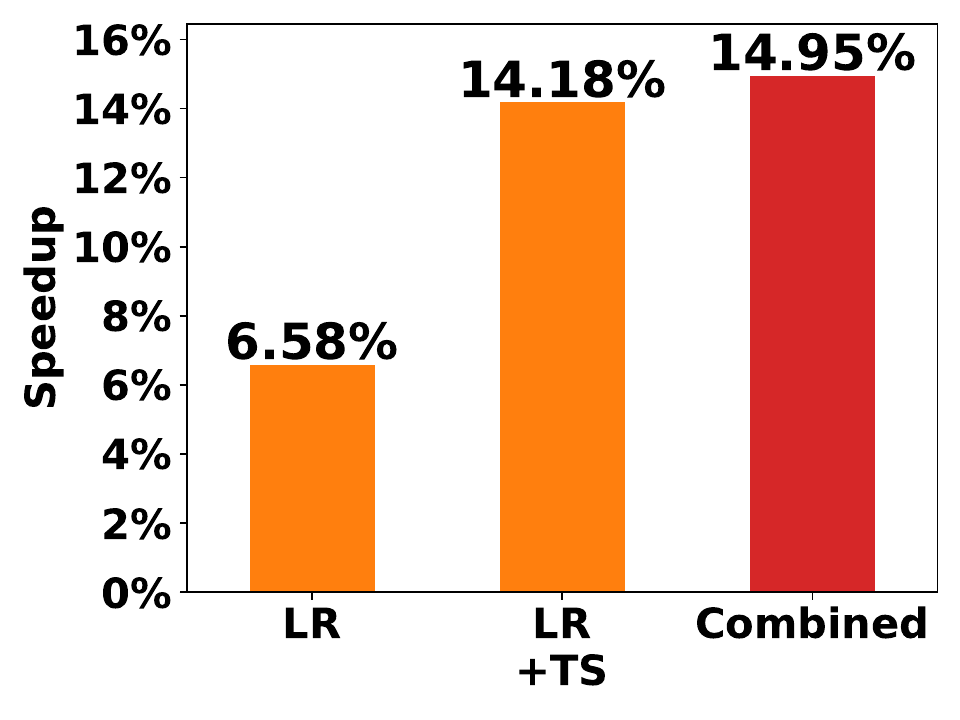}
    \caption{WRN18, CIFAR10}
    \label{fig:wideres-3}
  \end{subfigure}
    \begin{subfigure}[tb]{0.22\linewidth}
    \includegraphics[width=\linewidth]{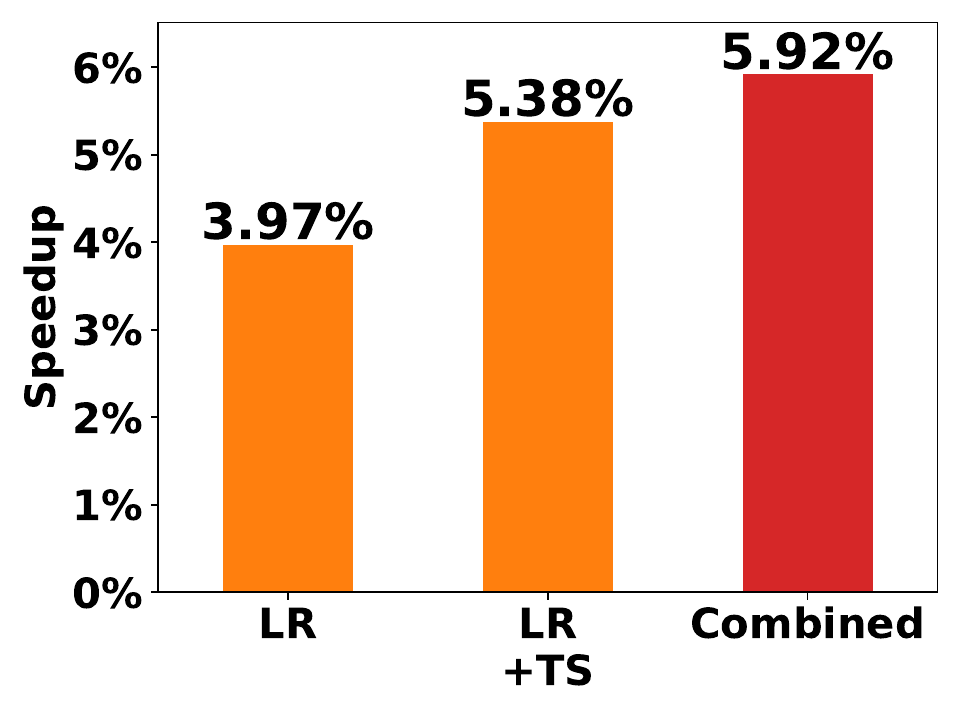}
    \caption{VGG19, ImageNet}
    \label{fig:vgg-3}
  \end{subfigure}
  \begin{subfigure}[tb]{0.22\linewidth}
    \includegraphics[width=\linewidth]{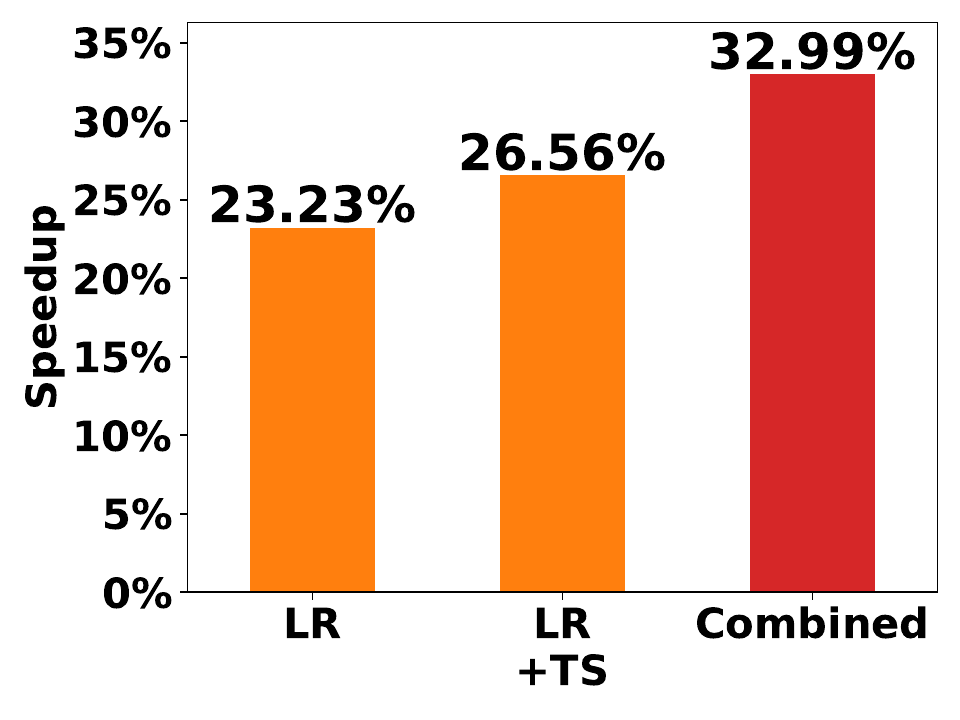}
    \caption{BERT\textsubscript{\scriptsize LARGE}, IMDb}
    \label{fig:bert-3}
  \end{subfigure}
  \smallskip 
  \caption{\name{} online phase performance improvement for Trio.}
  \label{fig:3pc-speedups}
\end{figure*}
\begin{figure*}[!h]
  \centering
  \begin{subfigure}[tb]{0.22\linewidth}
    \includegraphics[width=\linewidth]{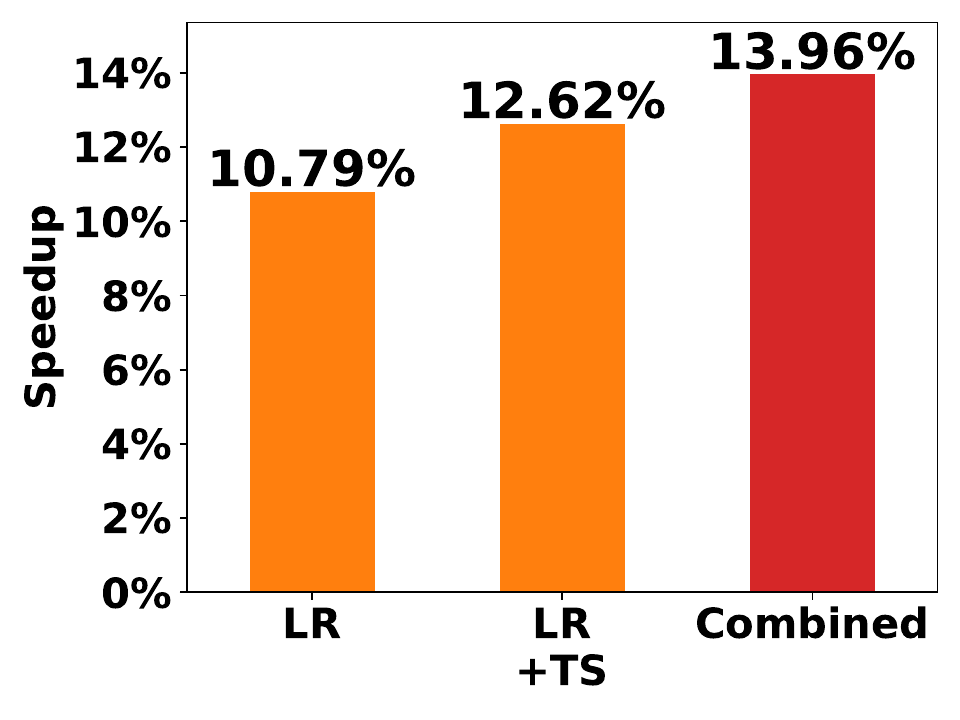}
    \caption{GCN, Cora}
    \label{fig:gcn-n}
  \end{subfigure}
  \begin{subfigure}[tb]{0.22\linewidth}
    \includegraphics[width=\linewidth]{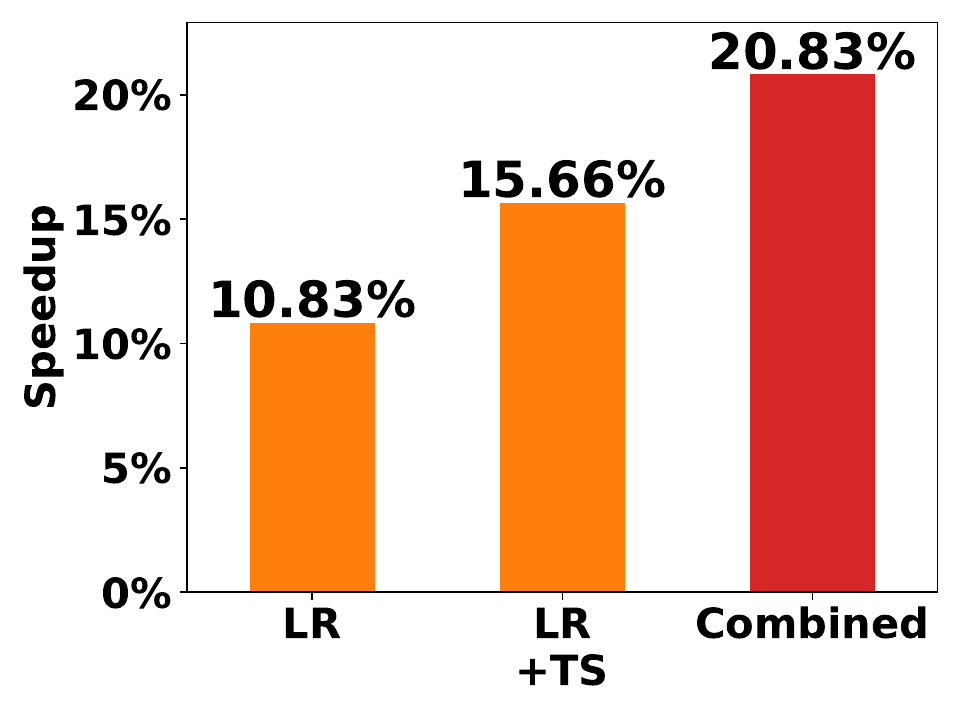}
    \caption{WRN18, CIFAR10}
    \label{fig:wideres-n}
  \end{subfigure}
    \begin{subfigure}[tb]{0.22\linewidth}
    \includegraphics[width=\linewidth]{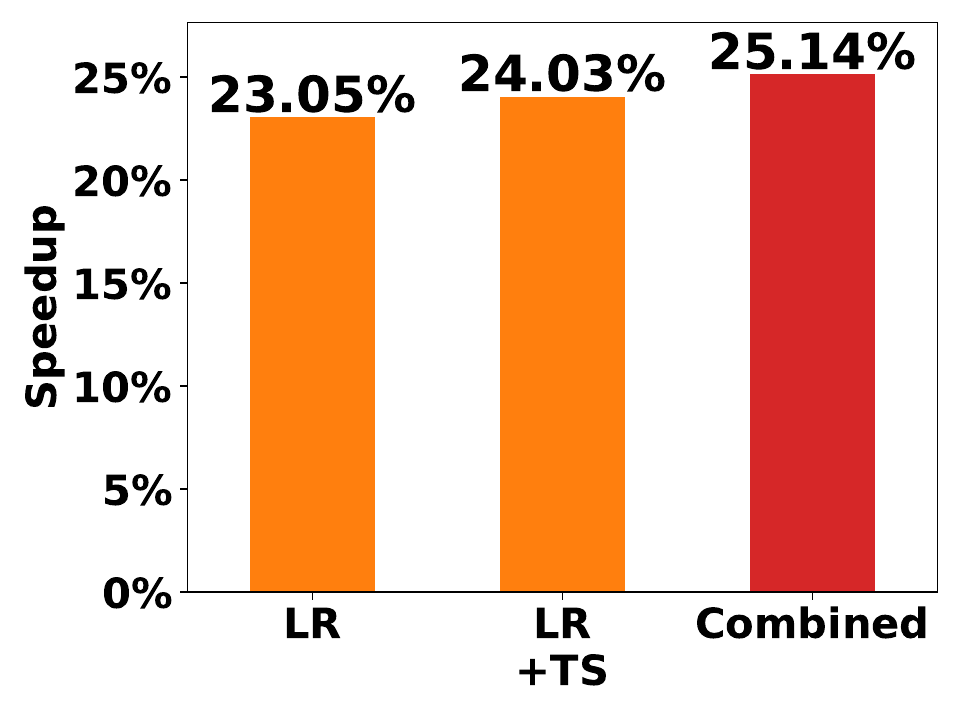}
    \caption{VGG19, ImageNet}
    \label{fig:vgg-n}
  \end{subfigure}
  \begin{subfigure}[tb]{0.22\linewidth}
    \includegraphics[width=\linewidth]{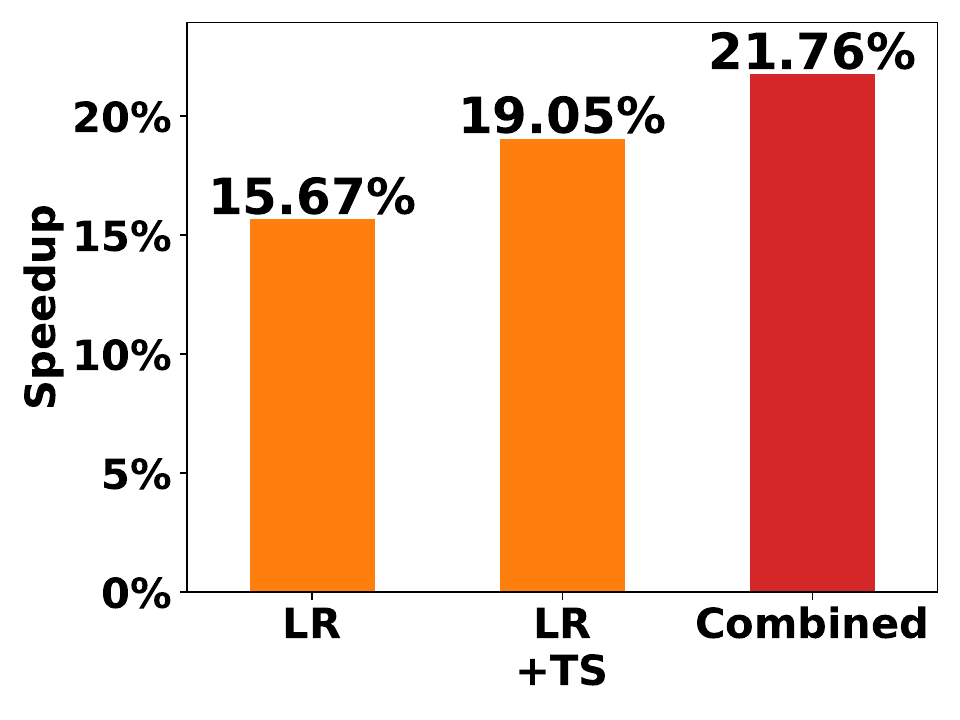}
    \caption{BERT\textsubscript{\scriptsize LARGE}, IMDb}
    \label{fig:bert-n}
  \end{subfigure}
  \smallskip 
  \caption{\name{} online phase performance improvement for \pc{n}.}
  \label{fig:npc-speedups}
\end{figure*}

\subsection{Model configurations}
\label{subsec:configuration}
In this subsection, we will discuss which layer we apply low-rank decomposition and truncation skipping for each model.

\para{Low-rank decomposition} If we apply low-rank decomposition to a Conv2d layer, the rank ratio (Section~\ref{sec:background}) is selected to be \textbf{0.5}. If we apply low-rank decomposition to an FC layer, the rank ratio will be \textbf{0.25}. These hyper parameters provided virtually no accuracy drop and have also been suggested in prior work~\cite{pufferfish}.

\para{Vision models} For both VGG and WideResNet, we adopt the low-rank decomposition policy outlined in~\cite{pufferfish}, as it ensures the best accuracy for these models. VGG19 consists of 16 Conv2d layers and three FC layers. Low-rank decomposition is applied to Conv2d layers ranging from  \textbf{the 10th to the 16th}, while the first nine Conv2d layers remain as full-rank convolutions. As for FC layers in VGG19, we apply low-rank decomposition to the first two FC layers. 
For WideResNet18, we apply low-rank decomposition exclusively to Conv2d layers from \textbf{the third to the 18th}, excluding any downsampling layers. All FC layers are kept as full-rank to maintain accuracy.

\para{GCN} Low-rank decomposition is applied to the matrix multiplications within the \graphconv{} operation. Specifically, both $\mW$ and $\mA$ are decomposed into smaller matrices, optimizing the computational efficiency while preserving the model's effectiveness for node classification.

\para{\bert{}} For \bert{}, we adopt the same low-rank decomposition policy for the Transformer architecture outlined in plaintext approach \cite{pufferfish}. \textbf{The very first Transformer layer is kept as a full-rank layer, and all other Transformer layers are low-rank layers.} For each low-rank Transformer layer, we use the rank ratio at 0.25 and decompose the weight matrices in the multi-head attention and the FC layers in the feed-forward networks (FFN). The embedding layer and the prediction layer are kept as full-rank.

\para{Truncation  Skipping} To ensure consistent evaluation, we conservatively enable truncation skipping for all low-rank layers across all models. While truncation skipping may impact model accuracy, our evaluations show that it maintains acceptable accuracy levels while optimizing performance.
%
With the model architectures and configurations outlined, we now turn our attention to discussing \name{}'s impact on the inference runtime of these models.

\subsection{Online phase speedups}
\label{subsec:online}
In this section, we analyze the online phase runtime speedup of \name{}. We calculate the runtime speedup using the equation $(\frac{T_{base}}{T_{ours}} - 1)\times100\%$, where $T_{base}$ is the baseline runtime measured using MPC on full rank matrices, and $T_{ours}$ is the runtime of MPC ML with our optimizations. 

\para{Trio protocol} Figure~\ref{fig:3pc-speedups} presents the online phase speedup achieved by \name{} across different models in the Trio protocol. The first bar (LR) shows how much speedup is acheived by basic low-rank without any of our optimizations on top of a full rank baseline. The remaining bars illustrate the contributions of truncation skipping (TS), and the combined effect with efficient linear layer concatenation (Combined).
Truncation skipping (TS) yields a significant improvement in speedup across all models, with WideResNet18 exhibiting the highest gain at approximately \pcent{7.5} (\pcent{6.58} → \pcent{14.18}). These results indicate that removing redundant truncation operations effectively reduces computational overhead, leading to improved efficiency.
Efficient linear layer concatenation further enhances performance, though its impact varies across models. The improvement from LR + TS → Combined is more pronounced in GCN and \bert{}, suggesting that models with extensive matrix multiplications benefit more from inter-layer optimizations, which is the current model trend.
The Combined approach consistently achieves the highest speedup, demonstrating that truncation skipping and efficient linear layer concatenation are complementary techniques that, when applied together, maximize performance improvements.
The observed speedup varies across models due to differences in the proportion of linear vs. nonlinear layers and model architecture.
GCN achieves the highest speedup, benefiting from a low proportion of nonlinear layers. Since GCNs rely heavily on matrix multiplications, low-rank decomposition and inter-layer optimizations have a substantial impact on reducing computation time.
WideResNet18 exhibits a moderate speedup, as truncation skipping and layer concatenation contribute to efficiency gains, but the presence of nonlinear layers such as ReLU limits overall improvements.
VGG19 shows the smallest speedup, as most of its computation lies in full-rank convolutional and nonlinear layers unaffected by \name{}. To maintain accuracy, low-rank decomposition is applied only from the 10th convolutional layer onward, limiting its coverage to 9 of 19 linear layers. This also reduces opportunities for truncation skipping and layer concatenation, further constraining the performance gains from LRD-MPC.
%

\bert{} achieves a moderate speedup, demonstrating that Transformer-based models benefit from low-rank decomposition and truncation skipping, though nonlinear components still constrain overall improvements. The effectiveness of LRD-MPC in \bert{} is further supported by the QUAD approximations \cite{mpcformer}, which reduce the proportion of nonlinear computations. 

\para{\pc{n} protocol} Figure~\ref{fig:npc-speedups} presents the online phase speedup achieved by \name{} in the \pc{n} protocol.
The impact of truncation skipping and efficient linear layer concatenation follow a similar trend as that in Trio.
%
%
Due to different implementation, VGG19 sees a higher relative speedup in \pc{n} compared to other models, whereas in Trio, it achieved the smallest improvement.
GCN and \bert{} continue to benefit from \name{} in both protocols, though the improvements are more pronounced in Trio. This is because linear layers account for a larger portion of the online phase runtime in full-rank model inference under Trio compared to \pc{n}. Specifically, GCN's linear layers contribute 81.9\% of the runtime in Trio versus 36.1\% in \pc{n}, while \bert{} sees a contribution of 73.2\% in Trio compared to 51.6\% in \pc{n}. Consequently, optimizations targeting linear layers have a greater impact in Trio.

\subsection{Speedups on different networks}
This section evaluates speedup of \name{} for different network setups, using the same equation as Section ~\ref{subsec:online}. LAN reports a low-latency setup where servers are connected with a 10 Gbps bandwidth. MAN models a network with 5 Gbps bandwidth and 5 ms round-trip latency, while WAN simulates conditions with the same bandwidth but a higher round-trip latency of 35 ms.

\begin{table}[h]
    \centering
    \caption{Speedup of \name{} on different network setups.}
    \begin{tabular}{ccccc}
        \toprule
        \textbf{Protocol} & \textbf{Model} &  \textbf{LAN}  & \textbf{MAN}  & \textbf{WAN}\\
        \midrule
        \multirow{4}{*}{\raisebox{-3.5ex}{Trio}} &
        GCN & 44.8\% & 44.45\% & 30.23\%\\
        \cmidrule{2-5}
        &WideResNet18 &  37.49\% & 28.06\% & 14.95\%\\
        \cmidrule{2-5}
        & VGG19 &  11.36\% & 10.92\% & 5.92\%\\
        \cmidrule{2-5}
        & BERT\textsubscript{\scriptsize LARGE} & 39.64\% & 37.91\% & 32.99\%\\

        \midrule

        \multirow{4}{*}{\raisebox{-3.5ex}{\pc{n}}} &
        GCN & 37.41\% & 36.01\% & 13.96\%\\
        \cmidrule{2-5}
        &WideResNet18 &  174.76\% & 125.64\% & 20.83\%\\
        \cmidrule{2-5}
        & VGG19 &  115.71\% & 113.96\% & 25.14\%\\
        \cmidrule{2-5}
        & BERT\textsubscript{\scriptsize LARGE} & 18.28\% & 13.53\% & 21.76\%\\
        \bottomrule
    \end{tabular}
    \label{tab:network}
\end{table}As shown in Table~\ref{tab:network}, across both MPC protocols (Trio and \pc{n}), \name{} consistently shows solid improvements across all networks, with gains becoming more notable under LAN setup. For example, under the \pc{n} protocol, WideResNet18 achieves a 174.76\% speedup in LAN, compared to 125.64\% in MAN and 20.83\% in WAN. This trend is expected, as computation becomes the dominant bottleneck when communication delays are reduced. These results confirm that \name{} provides strong performance improvements across a range of network setups. While speedups are the highest under low-latency conditions, \name{} maintains its advantage even as communication overhead increases, making it a practical and scalable optimization across deployment scenarios.

\subsection{Power consumption}
\label{subsec:eval-power}
In this section, we demonstrate the online phase power consumption savings of \name{}. In CrypTen, we run private inference on GPU and measure the GPU power consumption for both the full-rank model and our method \name{}. As shown in Table~\ref{tab:online_power}, \name{} reduces power consumption across multiple models by up to 52.2\%, demonstrating significant energy savings. At the data center scale, such reductions translate to substantial cost savings and improved sustainability for private inference workloads, making \name{} a practical solution for energy-efficient secure computation. 
\begin{table}[h]
    \centering
    \caption{\name{} online phase GPU power consumption saving (\%) for \pc{n}}
    \begin{tabular}{cc}
        \toprule
         \textbf{Model} & \textbf{Saving (\%)} \\
        \midrule
        GCN & 30.26\%\\
        \midrule
        WideResNet18 & 52.24\% \\
        \midrule
        VGG19 & 48.41\% \\
        \midrule BERT\textsubscript{\scriptsize LARGE} & 34.09\% \\
        \bottomrule
    \end{tabular}
    \label{tab:online_power}
\end{table}Power usage reduction is more pronounced than the reduction in runtime because, in MPC ML, linear layers dominate energy consumption. While activation functions may have longer runtime, most of activation runtime is spent on energy lightweight operations like communication and bitwise XOR and AND operations. Thus, \name{}, which significantly improves the efficiency of linear layers, can make a huge dent in energy usage.

Although equipment constraints and design choices in the PIGEON framework prevent direct energy logging for Trio, we expect its energy profile to align closely with that of the \pc{n} protocols.

\subsection{Offline phase speedups}
\label{subsec:offline}
This section evaluates the offline phase runtime speedup of \name{}.

\para{Trio protocol}
The offline phase in the Trio protocol is designed to minimize communication overhead. Specifically, $\party{1}$ generates all required random masks and sends all preprocessing materials to $\party{3}$ in a single communication round. Consequently, \name{}'s offline phase speedup stems solely from low-rank decomposition, as truncation skipping does not apply since no truncation occurs in offline phases, and efficient linear layer concatenation is irrelevant due to the absence of inter-layer communication.
Table~\ref{tab:offline_speedup} presents the \name{} offline phase speedup for various models in the Trio protocol, with GCN achieving highest improvement over \pcent{88}.


\begin{table}[h]
    \centering
    \begin{minipage}[t]{0.48\linewidth}
        \vspace{0pt}
        \centering
        \caption{\name{} offline phase speedup for Trio}
        \begin{tabular}{cc}
            \toprule
            \textbf{Model} & \textbf{Speedup} \\
            \midrule
            GCN & 88.27\%\\
            \midrule
            WideResNet18 & 15.38\% \\
            \midrule
            VGG19 & 5.28\% \\
            \midrule BERT\textsubscript{\scriptsize LARGE} & 35.69\% \\
            \bottomrule
        \end{tabular}
        \label{tab:offline_speedup}
    \end{minipage}
    \hfill
    \begin{minipage}[t]{0.48\linewidth}
        \vspace{0pt}
        \centering
        \caption{Reduction in Beaver triples generated for \pc{n} using \name{}}
        \begin{tabular}{cc}
            \toprule
            \textbf{Model} & \textbf{Reduction} \\
            \midrule
            GCN & 55.98\%\\
            \midrule
            WideResNet18 & 37.25\% \\
            \midrule
            VGG19 & 17.16\% \\
            \midrule BERT\textsubscript{\scriptsize LARGE} & 55.29\% \\
            \bottomrule
        \end{tabular}
        \label{tab:offline_beaver}
    \end{minipage}
\end{table}

The variation in speedup is driven by the proportion of linear vs. nonlinear layers in each model. GCN benefits the most because nonlinear layers constitute a relatively small portion of the model, allowing low-rank decomposition to substantially reduce computation time, leading to a high speedup. \bert{} sees a moderate speedup, as the QUAD approximations reduce the impact of nonlinear operations, making its computations more amenable to low-rank optimization. In contrast, VGG19 and WideResNet18 contain a higher proportion of nonlinear layers, such as ReLU and pooling layers, which remain unaffected by low-rank decomposition. As a result, the gains from optimizing linear layers translate to a smaller overall speedup in these models.

\para{\pc{n} protocol} Since CrypTen implements only the online phase, we cannot directly measure the offline phase speedup for \pc{n}. Instead, we report the reduction in the number of Beaver triples required to be generated for the linear layers when applying \name{} compared to the full-rank baseline. Table~\ref{tab:offline_beaver} shows that \name{} significantly decreases the number of required triples for linear layers across all models, directly minimizing the computational overhead of the offline phase. This is particularly important because the offline phase is often a bottleneck in MPC protocols due to the high cost of generating correlated randomness.

Overall, these results demonstrate that \name{} effectively reduces offline phase overhead, particularly for models where linear layers dominate computation, reinforcing its practicality for improving MPC efficiency in ML workloads.

\subsection{Model accuracy}
\label{subsec:accuracy}
To evaluate the impact of \name{} on model accuracy, we present inference results in Table~\ref{tab:accuracy}. \textbf{Low-rank} represents the baseline where low-rank decomposition is applied to the pre-trained model weights, without truncation skipping and efficient linear layer concatenation techniques, and \textbf{\name{}} includes all the optimizations we proposed in this paper.


\begin{table}[h]
    \centering
    \caption{\name{} and baseline model accuracy}
    \begin{tabular}{cccc}
        \toprule
        \textbf{Dataset} & \textbf{Model} &  \textbf{Low-rank}  & \textbf{\name{}} \\
        \midrule
        Cora & GCN & 83\% & 82.9\%\\
        \midrule
        CIFAR10  & WideResNet18 &  92.83\% & 92.83\%\\
        \midrule
        ImageNet & VGG19 &  56.3\% & 56.5\%\\
        \bottomrule
    \end{tabular}
    \label{tab:accuracy}
\end{table}
Our primary focus is the accuracy drop between the low-rank baseline and \name{}, as this isolates the impact of truncation skipping. Table~\ref{tab:accuracy} shows that all models maintain a minimal accuracy drop within \pcent{0.5}, including BERT\textsubscript{\scriptsize LARGE} (\pcent{0.32}), indicating that truncation skipping has a negligible impact.
BERT\textsubscript{\scriptsize LARGE} experiences a $>$\pcent{10} accuracy drop when transitioning from full-rank (\pcent{96}) to the low-rank baseline, an expected decline. Prior works~\cite{pufferfish, anotherfish, linformer} mitigate this by training low-rank models to recover full-rank accuracy, making it complementary to \name{}.


\section{Conclusion}
\label{sec:conclusion}
In this paper, we introduce \name{}, a suite of optimizations for efficient low-rank decomposition in MPC ML. Our approach hides and removes the additional overheads introduced by low-rank approximations in MPC ML through techniques such as truncation skipping and efficient linear layer concatenation. We demonstrate that \name{} improves inference runtime, power consumption, and offline phase costs by up to \pcent{33}, \pcent{52}, and \pcent{88}, respectively.


\bibliographystyle{ACM-Reference-Format}
\bibliography{references}

@misc{snoop1,
  title = { I2C bus monitor},
  howpublished = {\url{https://www.jupiteri.com/}},
  note = {Accessed: 2023-08-3}
}

@article{hashemi2022data,
      title={Data Leakage via Access Patterns of Sparse Features in Deep Learning-based Recommendation Systems}, 
      author={Hanieh Hashemi and Wenjie Xiong and Liu Ke and Kiwan Maeng and Murali Annavaram and G. Edward Suh and Hsien-Hsin S. Lee},
      year={2022},
      eprint={2212.06264},
      archivePrefix={arXiv},
      primaryClass={cs.CE}
}

@INPROCEEDINGS{specre,
  author={Kocher, Paul and Horn, Jann and Fogh, Anders and Genkin, Daniel and Gruss, Daniel and Haas, Werner and Hamburg, Mike and Lipp, Moritz and Mangard, Stefan and Prescher, Thomas and Schwarz, Michael and Yarom, Yuval},
  booktitle={2019 IEEE Symposium on Security and Privacy (SP)}, 
  title={Spectre Attacks: Exploiting Speculative Execution}, 
  year={2019},
  volume={},
  number={},
  pages={1-19},
  doi={10.1109/SP.2019.00002}}

@inproceedings{meltdown,
 author = {Moritz Lipp and Michael Schwarz and Daniel Gruss and Thomas Prescher and Werner Haas and Anders Fogh and Jann Horn and Stefan Mangard and Paul Kocher and Daniel Genkin and Yuval Yarom and Mike Hamburg},
 title = {Meltdown: Reading Kernel Memory from User Space},
 booktitle = {27th {USENIX} Security Symposium ({USENIX} Security 18)},
 year = {2018},
}

@INPROCEEDINGS{pufferfish,
  author={Hongyi Wang and Saurabh Agarwal and Dimitris Papailiopoulos},
  booktitle={The Fourth Annual Conference on Machine Learning and Systems}, 
  title={Pufferfish: Communication-efficient Models At No Extra Cost}, 
  year={2021}}

@inproceedings {wangCharacIspass,
author = {Yongqin Wang and G. Edward Suh and Wenjie Xiong and Benjamin Lefaudeux and Brian Knott and Murali Annavaram and Hsein-Hsin S. Lee},
booktitle = {2022 IEEE International Symposium on Performance Analysis of Systems and Software},
title = {Characterization of MPC-based Private Inference for Transformer-based Models},
year = {2022},
volume = {},
issn = {},
pages = {187-197},
doi = {10.1109/ISPASS55109.2022.00025},
url = {https://doi.ieeecomputersociety.org/10.1109/ISPASS55109.2022.00025},
publisher = {IEEE Computer Society},
address = {Los Alamitos, CA, USA},
month = {may}
}

@inproceedings {mpcpipe,
author = {Yongqin Wang and Rachit Rajat and Murali Annavaram},
booktitle = {2024 ACM Conference on Architectural Support for Programming Languages and Operating Systems},
title = {MPC-Pipe: An Efficient Pipeline Scheme for Semi-honest MPC Machine Learning},
year = {2024}
}

@INPROCEEDINGS{trio,
  author={Christopher Harth-Kitzerow and Ajith Suresh and Yongqin Wang and Hossein Yalame and Georg Carle, Murali Annavaram},
  title={High-Throughput Secure Multiparty Computation with an Honest Majority in Various Network Settings},
  year={2025},
  cdate={1704067200000},
  booktitle={25th Privacy Enhancing Technologies Symposium (PETS 2025)},
  volume={2025}
}

@inproceedings{crypten2020,
  author       = {Brian Knott and
                  Shobha Venkataraman and
                  Awni Y. Hannun and
                  Shubho Sengupta and
                  Mark Ibrahim and
                  Laurens van der Maaten},
  title        = {CrypTen: Secure Multi-Party Computation Meets Machine Learning},
  booktitle    = {Advances in Neural Information Processing Systems 34: Annual Conference
                  on Neural Information Processing Systems 2021, NeurIPS 2021, December
                  6-14, 2021, virtual},
  pages        = {4961--4973},
  year         = {2021}
}

@InProceedings{newprimitivempc,
author="Catrina, Octavian
and de Hoogh, Sebastiaan",
editor="Garay, Juan A.
and De Prisco, Roberto",
title="Improved Primitives for Secure Multiparty Integer Computation",
booktitle="Security and Cryptography for Networks",
year="2010",
publisher="Springer Berlin Heidelberg",
address="Berlin, Heidelberg",
pages="182--199",
isbn="978-3-642-15317-4"
}

@misc{curl2024,
      author = {Manuel B. Santos and Dimitris Mouris and Mehmet Ugurbil and Stanislaw Jarecki and José Reis and Shubho Sengupta and Miguel de Vega},
      title = {Curl: Private {LLMs} through Wavelet-Encoded Look-Up Tables},
      howpublished = {Cryptology ePrint Archive, Paper 2024/1127},
      year = {2024},
      note = {\url{https://eprint.iacr.org/2024/1127}},
      url = {https://eprint.iacr.org/2024/1127}
}

@inproceedings {prob_trunc_wrong,
	author = {Yun Li and Yufei Duan and Zhicong Huang and Cheng Hong and Chao Zhang and Yifan Song},
	title = {Efficient {3PC} for Binary Circuits with Application to {Maliciously-Secure} {DNN} Inference},
	booktitle = {32nd USENIX Security Symposium (USENIX Security 23)},
	year = {2023},
	isbn = {978-1-939133-37-3},
	address = {Anaheim, CA},
	pages = {5377--5394},
	url = {https://www.usenix.org/conference/usenixsecurity23/presentation/li-yun},
	publisher = {USENIX Association},
	month = aug
}

@inproceedings{asplos23character,
author = {Garimella, Karthik and Ghodsi, Zahra and Jha, Nandan Kumar and Garg, Siddharth and Reagen, Brandon},
title = {Characterizing and Optimizing End-to-End Systems for Private Inference},
year = {2023},
isbn = {9781450399180},
publisher = {Association for Computing Machinery},
address = {New York, NY, USA},
url = {https://doi.org/10.1145/3582016.3582065},
doi = {10.1145/3582016.3582065},
booktitle = {Proceedings of the 28th ACM International Conference on Architectural Support for Programming Languages and Operating Systems, Volume 3},
pages = {89–104},
numpages = {16},
location = {<conf-loc>, <city>Vancouver</city>, <state>BC</state>, <country>Canada</country>, </conf-loc>},
series = {ASPLOS 2023}
}

@misc{awsbandwidth,
  title = {Amazon EC2 instance network bandwidth},
  author = {Amazon AWS},
  howpublished = {\url{https://docs.aws.amazon.com/AWSEC2/latest/UserGuide/ec2-instance-network-bandwidth.html
}},
  note = {Accessed: 2024-08-20}
}

@misc{pigeon,
      author = {Christopher Harth-Kitzerow and Yongqin Wang and Rachit Rajat and Georg Carle and Murali Annavaram},
      title = {{PIGEON}: A Framework for Private Inference of Neural Networks},
      howpublished = {Cryptology {ePrint} Archive, Paper 2024/1371},
      year = {2024},
      url = {https://eprint.iacr.org/2024/1371}
}

@INPROCEEDINGS{wideresnet,
    author = {Sergey Zagoruyko and Nikos Komodakis},
    title = {Wide Residual Networks},
    booktitle = {BMVC},
    year = {2016}}

@article{vgg,
  author = {Simonyan, Karen and Zisserman, Andrew},
  bibsource = {dblp computer science bibliography, http://dblp.org},
  journal = {CoRR},
  title = {Very Deep Convolutional Networks for Large-Scale Image Recognition},
  url = {http://arxiv.org/abs/1409.1556},
  volume = {abs/1409.1556},
  year = 2014
}

@inproceedings{gcn,
title={Semi-Supervised Classification with Graph Convolutional Networks},
author={Thomas N. Kipf and Max Welling},
booktitle={International Conference on Learning Representations},
year={2017},
url={https://openreview.net/forum?id=SJU4ayYgl}
}

@inproceedings{bert,
    title = "{BERT}: Pre-training of Deep Bidirectional Transformers for Language Understanding",
    author = "Devlin, Jacob  and
      Chang, Ming-Wei  and
      Lee, Kenton  and
      Toutanova, Kristina",
    editor = "Burstein, Jill  and
      Doran, Christy  and
      Solorio, Thamar",
    booktitle = "Proceedings of the 2019 Conference of the North {A}merican Chapter of the Association for Computational Linguistics: Human Language Technologies, Volume 1 (Long and Short Papers)",
    month = jun,
    year = "2019",
    address = "Minneapolis, Minnesota",
    publisher = "Association for Computational Linguistics",
    url = "https://aclanthology.org/N19-1423/",
    doi = "10.18653/v1/N19-1423",
    pages = "4171--4186"
}

@inproceedings{cryptgpu,
  author     = {Sijun Tan and Brian Knott and Yuan Tian and David J. Wu},
  title      = {\textsc{CryptGPU}: Fast Privacy-Preserving Machine Learning on the GPU},
  booktitle  = {{IEEE} {S\&P}},
  year       = {2021}
}

@inproceedings{piranha,
  title={Piranha: A $\{$GPU$\}$ platform for secure computation},
  author={Watson, Jean-Luc and Wagh, Sameer and Popa, Raluca Ada},
  booktitle={31st USENIX Security Symposium (USENIX Security 22)},
  pages={827--844},
  year={2022}
}

@article{quantization1,
  title={Advances in the Neural Network Quantization: A Comprehensive Review},
  author={Anonymous},
  journal={Applied Sciences},
  volume={14},
  number={17},
  pages={7445},
  year={2024},
  publisher={MDPI}
}

@article{quantization2,
  title={A White Paper on Neural Network Quantization},
  author={Markus Nagel and Marios Fournarakis and Rana Ali Amjad and Yelysei Bondarenko and Mart van Baalen and Tijmen Blankevoort},
  journal={arXiv preprint arXiv:2106.08295},
  year={2021}
}

@misc{relubits,
      title={Approximating ReLU on a Reduced Ring for Efficient MPC-based Private Inference}, 
      author={Kiwan Maeng and G. Edward Suh},
      year={2023},
      eprint={2309.04875},
      archivePrefix={arXiv},
      primaryClass={cs.LG},
      url={https://arxiv.org/abs/2309.04875}, 
}

@article{mpcformer,
  title={{MPCFormer}: fast, performant and private Transformer inference with {MPC}},
  author={Li, Dacheng and Shao, Rulin and Wang, Hongyi and Guo, Han and Xing, Eric P. and Zhang, Hao},
  journal={arXiv preprint arXiv:2211.01452},
  year={2022},
  url={https://arxiv.org/abs/2211.01452},
  note={ICLR 2023 notable top 25\%}
}

@misc{reluskip,
      title={DeepReDuce: ReLU Reduction for Fast Private Inference}, 
      author={Nandan Kumar Jha and Zahra Ghodsi and Siddharth Garg and Brandon Reagen},
      year={2021},
      eprint={2103.01396},
      archivePrefix={arXiv},
      primaryClass={cs.LG},
      url={https://arxiv.org/abs/2103.01396}, 
}

@misc{cicra,
      title={Circa: Stochastic ReLUs for Private Deep Learning}, 
      author={Zahra Ghodsi and Nandan Kumar Jha and Brandon Reagen and Siddharth Garg},
      year={2021},
      eprint={2106.08475},
      archivePrefix={arXiv},
      primaryClass={cs.LG},
      url={https://arxiv.org/abs/2106.08475}, 
}

@misc{beavergen,
      author = {Deevashwer Rathee and Thomas Schneider and K.  K.  Shukla},
      title = {Improved Multiplication Triple Generation over Rings via {RLWE}-based {AHE}},
      howpublished = {Cryptology {ePrint} Archive, Paper 2019/577},
      year = {2019},
      doi = {10.1007/978-3-030-31578-8_19},
      url = {https://eprint.iacr.org/2019/577}
}

@misc{anotherfish,
      title={Cuttlefish: Low-Rank Model Training without All the Tuning}, 
      author={Hongyi Wang and Saurabh Agarwal and Pongsakorn U-chupala and Yoshiki Tanaka and Eric P. Xing and Dimitris Papailiopoulos},
      year={2023},
      eprint={2305.02538},
      archivePrefix={arXiv},
      primaryClass={cs.LG},
      url={https://arxiv.org/abs/2305.02538}, 
}

@inproceedings{aby3,
  title={ABY3: A mixed protocol framework for machine learning},
  author={Mohassel, Payman and Rindal, Peter},
  booktitle={Proceedings of the 2018 ACM SIGSAC conference on computer and communications security},
  pages={35--52},
  year={2018}
}

@inproceedings{astra,
  title={ASTRA: high throughput 3pc over rings with application to secure prediction},
  author={Chaudhari, Harsh and Choudhury, Ashish and Patra, Arpita and Suresh, Ajith},
  booktitle={Proceedings of the 2019 ACM SIGSAC Conference on Cloud Computing Security Workshop},
  pages={81--92},
  year={2019}
}

@misc{linformer,
      title={Linformer: Self-Attention with Linear Complexity}, 
      author={Sinong Wang and Belinda Z. Li and Madian Khabsa and Han Fang and Hao Ma},
      year={2020},
      eprint={2006.04768},
      archivePrefix={arXiv},
      primaryClass={cs.LG},
      url={https://arxiv.org/abs/2006.04768}, 
}

@misc{compacttag,
      author = {Yongqin Wang and Pratik Sarkar and Nishat Koti and Arpita Patra and Murali Annavaram},
      title = {{CompactTag}: Minimizing Computation Overheads in Actively-Secure {MPC} for Deep Neural Networks},
      howpublished = {Cryptology {ePrint} Archive, Paper 2023/1729},
      year = {2023},
      url = {https://eprint.iacr.org/2023/1729}
}

@INPROCEEDINGS{yao,
  author={Yao, Andrew Chi-Chih},
  booktitle={27th Annual Symposium on Foundations of Computer Science (sfcs 1986)}, 
  title={How to generate and exchange secrets}, 
  year={1986},
  volume={},
  number={},
  pages={162-167},
  doi={10.1109/SFCS.1986.25}}

@INPROCEEDINGS{secureml,
  author={Mohassel, Payman and Zhang, Yupeng},
  booktitle={2017 IEEE Symposium on Security and Privacy (SP)}, 
  title={SecureML: A System for Scalable Privacy-Preserving Machine Learning}, 
  year={2017},
  volume={},
  number={},
  pages={19-38},
  doi={10.1109/SP.2017.12}}

@inproceedings {fantasticfour,
	author = {Anders Dalskov and Daniel Escudero and Marcel Keller},
	title = {Fantastic Four: {Honest-Majority} {Four-Party} Secure Computation With Malicious Security},
	booktitle = {30th USENIX Security Symposium (USENIX Security 21)},
	year = {2021},
	isbn = {978-1-939133-24-3},
	pages = {2183--2200},
	url = {https://www.usenix.org/conference/usenixsecurity21/presentation/dalskov},
	publisher = {USENIX Association},
	month = aug
}

@inproceedings {swift,
author = {Nishat Koti and Mahak Pancholi and Arpita Patra and Ajith Suresh},
title = {{SWIFT}: Super-fast and Robust {Privacy-Preserving} Machine Learning},
booktitle = {30th USENIX Security Symposium (USENIX Security 21)},
year = {2021},
isbn = {978-1-939133-24-3},
pages = {2651--2668},
url = {https://www.usenix.org/conference/usenixsecurity21/presentation/koti},
publisher = {USENIX Association},
month = aug
}

@inproceedings{tetrad,
  author       = {Nishat Koti and
                  Arpita Patra and
                  Rahul Rachuri and
                  Ajith Suresh},
  title        = {Tetrad: Actively Secure 4PC for Secure Training and Inference},
  booktitle    = {29th Annual Network and Distributed System Security Symposium, {NDSS}
                  2022, San Diego, California, USA, April 24-28, 2022},
  publisher    = {The Internet Society},
  year         = {2022},
  url          = {https://www.ndss-symposium.org/ndss-paper/auto-draft-202/},
  bibsource    = {dblp computer science bibliography, https://dblp.org}
}

@inbook{pytorch,
author = {Paszke, Adam and Gross, Sam and Massa, Francisco and Lerer, Adam and Bradbury, James and Chanan, Gregory and Killeen, Trevor and Lin, Zeming and Gimelshein, Natalia and Antiga, Luca and Desmaison, Alban and K\"{o}pf, Andreas and Yang, Edward and DeVito, Zach and Raison, Martin and Tejani, Alykhan and Chilamkurthy, Sasank and Steiner, Benoit and Fang, Lu and Bai, Junjie and Chintala, Soumith},
title = {PyTorch: an imperative style, high-performance deep learning library},
year = {2019},
publisher = {Curran Associates Inc.},
address = {Red Hook, NY, USA},
booktitle = {Proceedings of the 33rd International Conference on Neural Information Processing Systems},
articleno = {721},
numpages = {12}
}

@misc{cuda, title={Cuda Libraries Documentation}, url={https://docs.nvidia.com/cuda-libraries/index.html}, journal={NVIDIA Developer Documentation}, author={n.d., NVIDIA Corporation.}}

@misc{cutlass,
  author={{NVIDIA Corporation. n.d}},
  title        = {CUTLASS: CUDA Templates for Linear Algebra Subroutines},
  howpublished ={\url{https://github.com/NVIDIA/cutlass}},
  journal={NVIDIA Developer Documentation}, 
}

@inproceedings{hummingbird,
 author = {Maeng, Kiwan and Suh, G. Edward},
 booktitle = {Proceedings of Machine Learning and Systems},
 editor = {P. Gibbons and G. Pekhimenko and C. De Sa},
 pages = {128--147},
 title = {Accelerating ReLU for MPC-Based Private Inference with a Communication-Efficient Sign Estimation},
 url = {https://proceedings.mlsys.org/paper_files/paper/2024/file/4e3157021c5f833bb2204081f1dda573-Paper-Conference.pdf},
 volume = {6},
 year = {2024}
}

@article{securenn,
  title={{S}ecure{NN}: 3-{P}arty {S}ecure {C}omputation for {N}eural {N}etwork {T}raining},
  author={Wagh, Sameer and Gupta, Divya and Chandran, Nishanth},
  journal={Proceedings on Privacy Enhancing Technologies},
  year={2019}
}

@inproceedings{cryptflow2,
author = {Rathee, Deevashwer and Rathee, Mayank and Kumar, Nishant and Chandran, Nishanth and Gupta, Divya and Rastogi, Aseem and Sharma, Rahul},
title = {CrypTFlow2: Practical 2-Party Secure Inference},
year = {2020},
isbn = {9781450370899},
publisher = {Association for Computing Machinery},
address = {New York, NY, USA},
url = {https://doi-org.libproxy2.usc.edu/10.1145/3372297.3417274},
doi = {10.1145/3372297.3417274},
booktitle = {Proceedings of the 2020 ACM SIGSAC Conference on Computer and Communications Security},
pages = {325–342},
numpages = {18},
location = {Virtual Event, USA},
series = {CCS '20}
}

@article{Falcon,
  title={Falcon: Honest-Majority Maliciously Secure Framework for Private Deep Learning},
  author={Sameer Wagh and Shruti Tople and Fabrice Benhamouda and Eyal Kushilevitz and Prateek Mittal and Tal Rabin},
  journal={Proceedings on Privacy Enhancing Technologies},
  year={2020},
  volume={2021},
  pages={188 - 208},
  url={https://api.semanticscholar.org/CorpusID:214802385}
}

@InProceedings{deepreduce,
  title = 	 {DeepReDuce: ReLU Reduction for Fast Private Inference},
  author =       {Jha, Nandan Kumar and Ghodsi, Zahra and Garg, Siddharth and Reagen, Brandon},
  booktitle = 	 {Proceedings of the 38th International Conference on Machine Learning},
  pages = 	 {4839--4849},
  year = 	 {2021},
  editor = 	 {Meila, Marina and Zhang, Tong},
  volume = 	 {139},
  series = 	 {Proceedings of Machine Learning Research},
  month = 	 {18--24 Jul},
  publisher =    {PMLR},
  url = 	 {https://proceedings.mlr.press/v139/jha21a.html}
}

@InProceedings{snl,
  title = 	 {Selective Network Linearization for Efficient Private Inference},
  author =       {Cho, Minsu and Joshi, Ameya and Reagen, Brandon and Garg, Siddharth and Hegde, Chinmay},
  booktitle = 	 {Proceedings of the 39th International Conference on Machine Learning},
  pages = 	 {3947--3961},
  year = 	 {2022},
  editor = 	 {Chaudhuri, Kamalika and Jegelka, Stefanie and Song, Le and Szepesvari, Csaba and Niu, Gang and Sabato, Sivan},
  volume = 	 {162},
  series = 	 {Proceedings of Machine Learning Research},
  month = 	 {17--23 Jul},
  publisher =    {PMLR},
  url = 	 {https://proceedings.mlr.press/v162/cho22a.html}
}

@inproceedings{
kundu,
title={Learning to Linearize Deep Neural Networks  for Secure and Efficient Private Inference},
author={Souvik Kundu and Shunlin Lu and Yuke Zhang and Jacqueline Tiffany Liu and Peter Anthony Beerel},
booktitle={The Eleventh International Conference on Learning Representations },
year={2023},
url={https://openreview.net/forum?id=BGF9IeDfmlH}
}

@inproceedings{lowranklth,
 author = {Schotth\"{o}fer, Steffen and Zangrando, Emanuele and Kusch, Jonas and Ceruti, Gianluca and Tudisco, Francesco},
 booktitle = {Advances in Neural Information Processing Systems},
 editor = {S. Koyejo and S. Mohamed and A. Agarwal and D. Belgrave and K. Cho and A. Oh},
 pages = {20051--20063},
 publisher = {Curran Associates, Inc.},
 title = {Low-rank lottery tickets: finding efficient low-rank neural networks via matrix differential equations},
 url = {https://proceedings.neurips.cc/paper_files/paper/2022/file/7e98b00eeafcdaeb0c5661fb9355be3a-Paper-Conference.pdf},
 volume = {35},
 year = {2022}
}

@inproceedings{
lora,
title={Lo{RA}: Low-Rank Adaptation of Large Language Models},
author={Edward J Hu and Yelong Shen and Phillip Wallis and Zeyuan Allen-Zhu and Yuanzhi Li and Shean Wang and Lu Wang and Weizhu Chen},
booktitle={International Conference on Learning Representations},
year={2022},
url={https://openreview.net/forum?id=nZeVKeeFYf9}
}

@inproceedings{lowrankproperty,
  title={Restructuring of deep neural network acoustic models with singular value decomposition.},
  author={Xue, Jian and Li, Jinyu and Gong, Yifan},
  booktitle={Interspeech},
  pages={2365--2369},
  year={2013}
}

\end{document}